\documentclass[fleqn,usenatbib]{mnras}

\usepackage{newtxtext,newtxmath}
\usepackage[T1]{fontenc}

\DeclareRobustCommand{\VAN}[3]{#2}
\let\VANthebibliography\thebibliography
\def\thebibliography{\DeclareRobustCommand{\VAN}[3]{##3}\VANthebibliography}

\usepackage{caption}
\usepackage{subcaption} 
\usepackage{graphicx}	% Including figure files
\usepackage{amsmath}	% Advanced maths commands
\usepackage{multicol}
\usepackage{orcidlink} % add orcid 
\usepackage{longtable} % for long tables
\usepackage[noabbrev,capitalize]{cleveref} % to avoid have to write Equation \ref{} or section \ref
\usepackage{booktabs}
\usepackage{here}
\usepackage{xspace}
\newcommand{\xmm}{\textit{XMM--Newton}\xspace}
\newcommand{\swift}{\textit{Swift}\xspace}
\newcommand{\angstrom}{\,\textup{\AA}}
\usepackage{pdflscape}
\usepackage{xcolor}
\usepackage{tikz}
\usepackage{pgf} 
\usepackage{amsmath}
\usetikzlibrary{positioning}
\usetikzlibrary{arrows}
\usetikzlibrary{fit}
\usetikzlibrary{shapes.geometric}
\usepackage{afterpage}

\title[X-ray power spectra of AGN from \textit{XMM} and \textit{Swift}]{The variability of active galaxies: I. Broad-band noise X-ray power spectra from \textit{XMM--Newton} and \textit{Swift}}

\author[Lefkir et al.]{
Mehdy Lefkir\,\orcidlink{0000-0002-6972-2429}$^{1}$\thanks{E-mail: ml556@leicester.ac.uk},
Simon Vaughan\,\orcidlink{0000-0003-4808-092X}$^{1}$,
Mike Goad\,\orcidlink{0000-0002-2908-7360}$^{1}$,
Daniela Huppenkothen\,\orcidlink{0000-0002-1169-7486}$^{2}$,
Phil Uttley\,\orcidlink{0000-0001-9355-961X}$^{2}$
\\
$^{1}$School of Physics and Astronomy, University of Leicester, Leicester, LE1 7RH, UK\\
$^{2}$Anton Pannekoek Institute for Astronomy, University of Amsterdam, Science PArk~904, NL-1098 XH Amsterdam, The Netherlands}
\date{Accepted XXX. Received YYY; in original form ZZZ}

\pubyear{2025}

\begin{document}
\label{firstpage}
\pagerange{\pageref{firstpage}--\pageref{lastpage}}
\maketitle

\begin{abstract}
Accreting supermassive black holes at the centres of galaxies are the engine of active galactic nuclei (AGN). X-ray light curves of unabsorbed AGN show dramatic random variability on timescales ranging from seconds to years. The power spectrum of the fluctuations is usually well-modelled with a power law that decays as $1/f$ at low frequencies, and which bends to $1/f^{2-3}$ at high frequencies. The timescale associated with the bend correlates well with the mass of the black hole and may also correlate with bolometric luminosity in the `X-ray variability plane'. Because AGN light curves are usually irregularly sampled, the estimation of AGN power spectra is challenging. In a previous paper, we introduced a new method to estimate the parameters of bending power law power spectra from AGN light curves. We apply this method to a sample of 56 variable and unabsorbed AGN, observed with \textit{XMM-Newton} and \textit{Swift} in the $0.3-1.5$\,keV band over the past two decades. We obtain estimates of the bends in 50 sources, which is the largest sample of X-ray bends in the soft band. We also find that the high-frequency power spectrum is often steeper than 2. We update the X-ray variability plane with new bend timescale measurements spanning from 7 min to 62 days. We report the detections of low-frequency bends in the power spectra of five AGN, three of which are previously unpublished: 1H~1934-063, Mkn~766 and Mkn~279.
\end{abstract}


%%%%%%%%%%%%%%%%%%%%%%%%%%%%%%%%%%%%%%%%%%%%%%%%%%
\defcitealias{2025MNRAS.539.1775L}{L25}

%%%%%%%%%%%%%%%%% BODY OF PAPER %%%%%%%%%%%%%%%%%%

\section{Introduction}

Active galaxies or Active Galactic Nuclei (AGN) are bright sources located in a compact region at the centres of galaxies \citep{2017A&ARv..25....2P}. Powered by the accretion of matter onto a supermassive black hole, AGN emit at all wavelengths and each waveband probes a different region of the AGN \citep{1984ARA&A..22..471R}. Matter is accreted via a disc; in the $\alpha$ prescription the disc is generally thought to be geometrically flat and optically thick \citep{1973A&A....24..337S,1973blho.conf..343N}. Gas particles in the disc are spiralling inwards by losing angular momentum, collisions within the disc generate heat which is radiated away following a multi-temperature black body peaking in the ultraviolet \citep{1981ARA&A..19..137P}. X-rays originate from innermost region of the AGN, they are thought to be produced by the up-scattering of disc photons in a hot plasma called the corona  \citep{1991ApJ...380L..51H,1993ApJ...413..507H}. The nature and geometry of the corona is still a mystery, however recent studies with X-ray polarimetry have started to provide constraints on the geometry \citep[][]{2024Galax..12...20K,2024ApJ...974..101S,2023MNRAS.523.4468G,2022MNRAS.516.5907M}. Due to the compact size  and the distance of the AGN, spatially resolving the central region is challenging in almost all sources. However, physical and geometrical properties of the AGN can be inferred using spectral and timing information. 

\begin{figure*}
\centering
\begin{subfigure}[b]{\textwidth}
    \includegraphics[width=\textwidth]{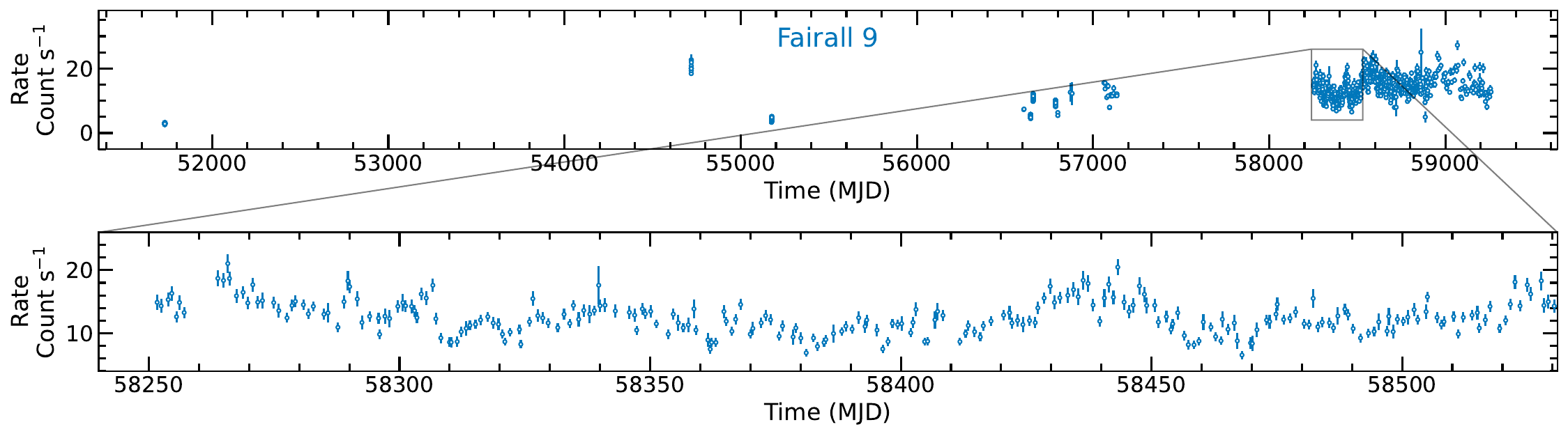}
    \caption{Combined \xmm{} and \swift{} light curve of Fairall~9 over 22 years in the $0.3-1.5$\,keV energy band. The inset shows a portion of intense \swift{} monitoring designed for reverberation mapping studies \citep{2020MNRAS.498.5399H}.}
    \label{fig:fairall_9_lc}
\end{subfigure}
\hfill\\
\begin{subfigure}[b]{\textwidth}
    \includegraphics[width=\textwidth]{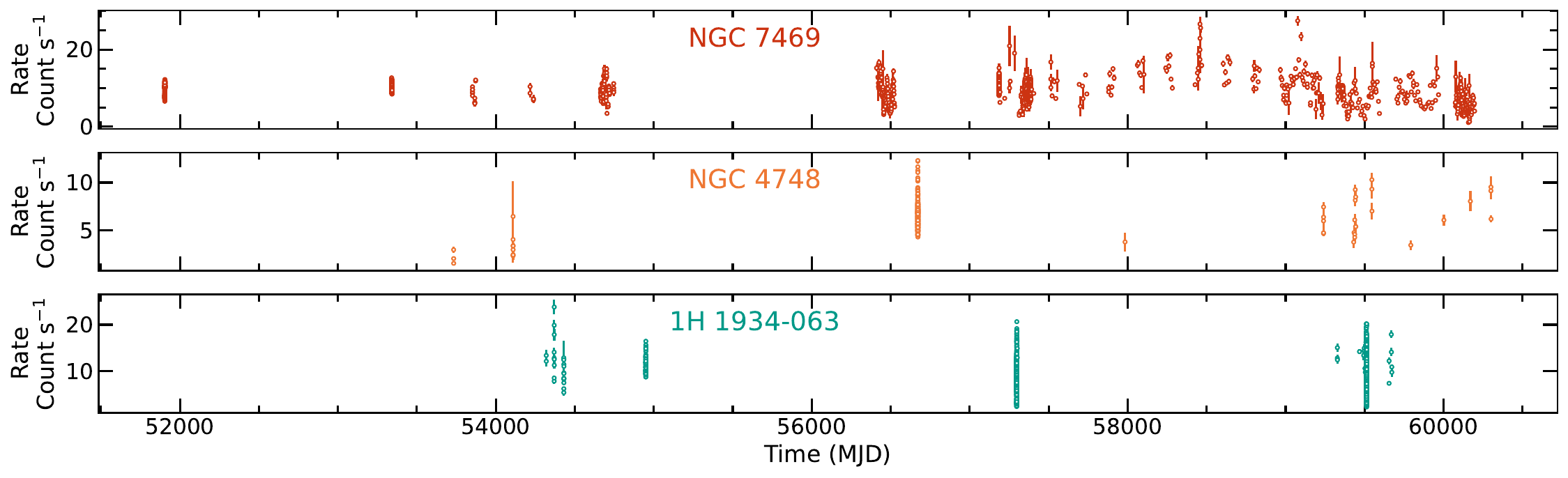}
    \caption{Combined \xmm{} and \swift{} light curves of NGC~7469, NGC~4748 and 1H~1934-063 in the $0.3-1.5$\,keV energy band.}
    \label{fig:full_lc}
\end{subfigure}\hfill\\
\begin{subfigure}[t]{0.59\textwidth}
    \includegraphics[width=\textwidth]{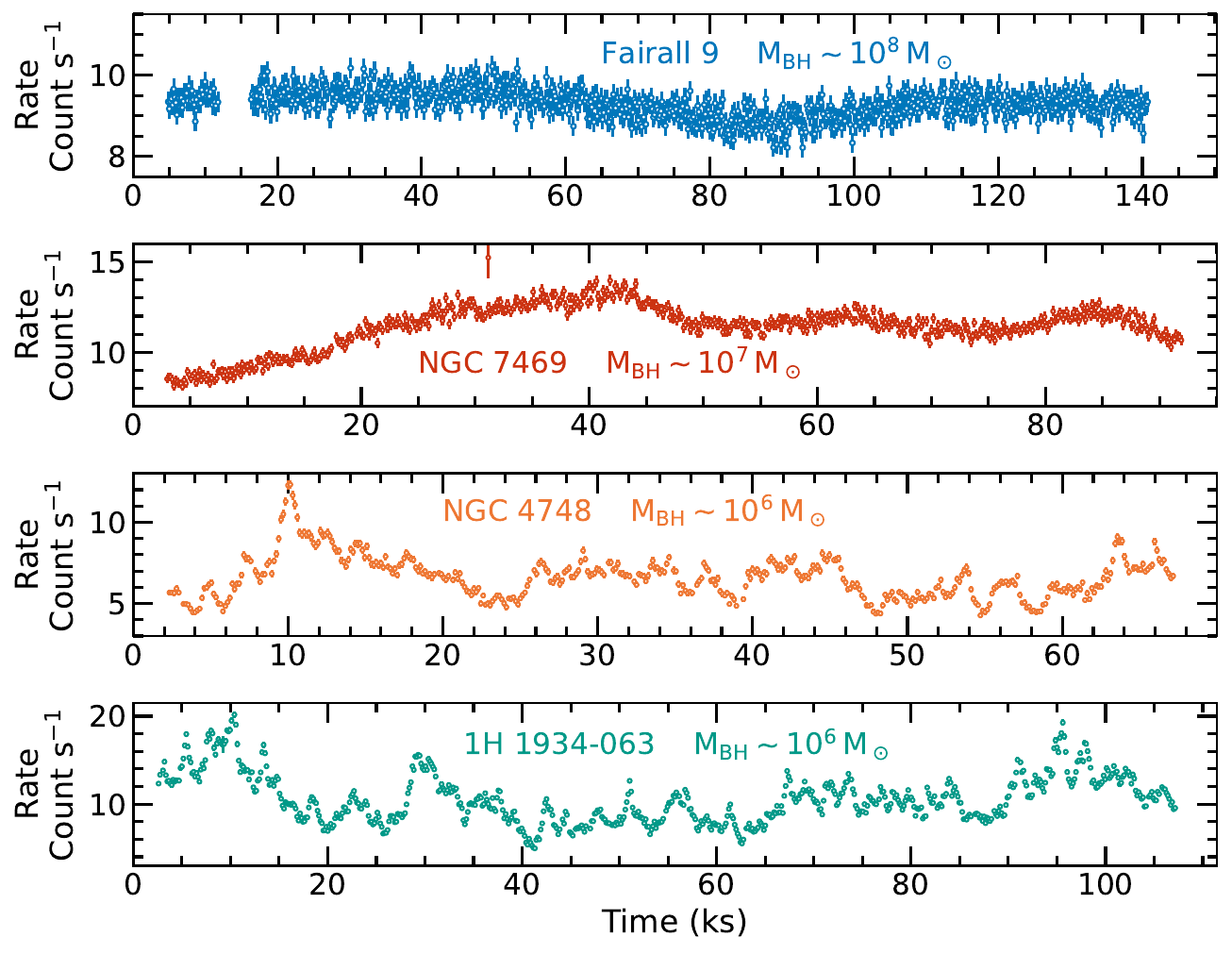}
    \caption{Segments of \xmm{} light curves in the $0.3-1.5$ keV band for Fairall~9, NGC~7469, NGC~4748 and 1H~1934-063 (from top to bottom). The text in each subfigure indicates the mass of the black hole. We see that within segments of $90-140$\,ks the variations become faster with larger amplitude when the mass decreases. The light curve of NGC~7469 also presents a \swift{} datapoint which appears as an outlier with respect to the \xmm{} light curve after cross-calibration. }
    \label{fig:xmm_lc}
\end{subfigure}\hfill
\begin{subfigure}[b]{0.4\textwidth}
    \includegraphics[width=\textwidth]{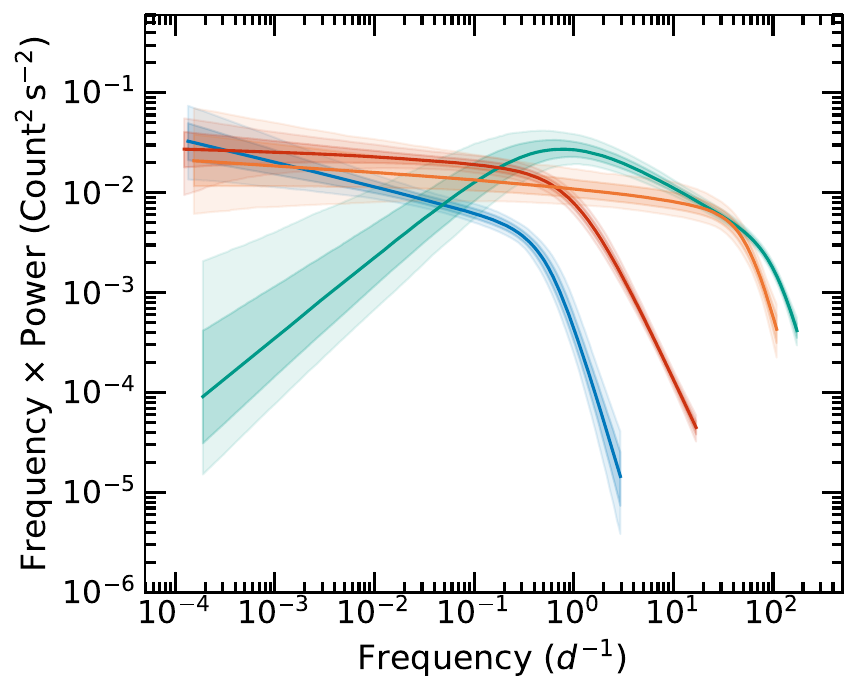}
    \caption{Posterior power spectral densities for Fairall~9, NGC~4748, 1H~1934-063 and NGC~7469 estimated in this work. We observe that the bend frequency moves towards low frequencies as the mass increases. The power spectra have different slopes at low and high frequencies. 1H~1934-063 shows strong evidence for two bends in its power spectrum.}
    \label{fig:powerspectra}
\end{subfigure}
        
\caption{Light curves and power spectral densities for Fairall~9, NGC~7469, NGC~4748 and 1H~1934-063. The sources are identified by a similar colour in all panels.}
\label{fig:lc}
\end{figure*}

One of the most common features of active galaxies is the flux variations observed from radio to $\gamma$-rays \citep{Paolillo2025,1997ARA&A..35..445U,1995ARA&A..33..163W,1981MNRAS.194..987M}. These variations are random often with a large amplitude in the X-rays over the timescales of years down to seconds. Figures~\ref{fig:fairall_9_lc}, \ref{fig:full_lc} and \ref{fig:xmm_lc} show examples of light curves for different sources and timescales. Random fluctuations in the mass accretion rate propagating inwards can explain the X-ray variability \citep{1997MNRAS.292..679L,2006MNRAS.367..801A}. These fluctuations originate from turbulence in the disc generated through magneto-rotational instability (MRI)  \citep{2020MNRAS.496.3808B,2016ApJ...826...40H,2004MNRAS.348..111K}. As discussed by \cite{2024MNRAS.530.4850H}, the variable UV/optical emission from the disc could drive the slow variations in the X-rays but would be uncorrelated with the fast X-ray variability generated in the corona. 

The fluctuations observed in X-ray light curves are often studied using the power spectrum, which quantifies the contribution of each frequency ($1/$timescale) to the total variance \citep{1989ASIC..262...27V,2003MNRAS.345.1271V}. AGN power spectra are dominated by broad-band noise (a smooth spectrum spanning a wide range of frequencies without narrow features) and are often well-described by power laws of the form $1/f^\alpha$, $\alpha \sim 0-1$, at low frequencies \citep{1987Natur.325..694L,1993MNRAS.265..664G,1993MNRAS.261..612P,2003ApJ...593...96M}. At higher frequencies, AGN light curves show variability akin to red noise processes as the power spectral density breaks to a steep power law $\alpha>2$ \citep[][]{1999ApJ...514..682E}. In \cref{fig:powerspectra}, we show the power spectra of various AGN. The timescale associated with the break is one of the few characteristic timescales observed and correlates well with the mass of the black hole \citep{2004MNRAS.348..783M,2006Natur.444..730M} - down to stellar-mass black holes observed in black hole X-ray binaries (BHXRBs). AGN and BHXRBs share emission and variability properties that hint at common physical processes in a black hole unification scheme \citep{2007arXiv0706.3838F,2006A&A...456..439K}. \cite{2006Natur.444..730M} defined an X-ray variability plane relating the bend timescale, the mass of the black hole and the bolometric luminosity. The normalised excess variance - which is related to the integral of the power spectrum \citep{1997ApJ...476...70N,2003MNRAS.339.1237V} - also correlates well with black hole mass \citep{2014ApJ...781..105L,2012A&A...542A..83P,2004MNRAS.348..207P}. The relation between excess variance and black hole mass is typical of the general AGN population, as is it also observed in high-redshift AGN \citep[][]{2023A&A...673A..68P,2017MNRAS.471.4398P}. The relationship between the break timescale and black hole mass was extended to other variable accreting compact objects such as white dwarfs and young stellar objects and also appears in the optical \citep[e.g.][]{2015SciA....1E0686S,2021Sci...373..789B}. 

AGN variability is usually considered stationary, the power spectrum does not appear to vary significantly with time \citep[][]{2003ApJ...593...96M,2011MNRAS.413.2489V} with some exceptions \citep[e.g.][]{2019MNRAS.482.2088A}. On the other hand, BHXRBs evolve over the course of days to weeks through various `states' which show dramatically different power spectral shapes \citep{2006csxs.book..157M,2015MNRAS.448.3339H}. In the soft state, the thermal emission from the disc dominates and the power spectrum is usually simple and similar to typical AGN power spectra \citep{2005MNRAS.363..586U,2007MNRAS.378..649S}. In the hard state, the Compton upscattered X-ray emission dominates the energy spectrum and the power spectrum is usually more complex, showing possibly multiple broad and narrow features, and often a break at low-frequencies. BHXRBs power spectra are usually modelled with a mixture of Lorentzians. As AGN are orders of magnitude more massive than BHXRBs, a timescale of a few hours in BHXRBs corresponds to centuries for most AGN. Ark~564 and IRAS~13224-3809 are the only two AGN \citep{2007MNRAS.382..985M,2019MNRAS.482.2088A} with evidence for a double-bending power spectrum making these sources analogues of BHXRBs in their hard state. BHXRBs also often show quasi-periodic oscillations (QPOs) \citep{2019NewAR..8501524I}, these are very uncommon for AGN except for a few sources \citep[see e.g.][]{2025Natur.638..370M,2016AN....337..417A}.

The power spectrum is usually estimated directly using a periodogram analysis \citep{2023hxga.book..141B,1989ASIC..262...27V} with the requirement that the light curve must be regularly sampled. Periodograms are biased by leakage and aliasing \citep{2002MNRAS.332..231U}. Monte Carlo simulations and linear interpolation were proposed by \cite{2002MNRAS.332..231U} to model the biases and alleviate the requirement on the sampling, however this approach is computationally expensive and the likelihood function was not properly defined. \cite{2014ApJ...788...33K} introduced a time domain approach named CARMA to model the power spectrum of irregularly sampled time series. This approach is based on Gaussian Process (GP) regression, it has a well-defined likelihood, it accounts for uncertainties in the time series and is immune to leakage and aliasing. The power spectrum of the CARMA process is expressed as a sum of modified Lorentzians. Estimating and interpreting the parameters of the CARMA model is usually challenging in the context of AGN variability as one may need a large number of modified Lorentzians to accurately reproduce the typical bending power law with arbitrary slopes. In \cite{2025MNRAS.539.1775L}, hereafter \citetalias{2025MNRAS.539.1775L}, we introduced \texttt{PIORAN} an alternative GP method which approximates a bending power law power spectrum in terms of  basis functions. 

In this paper, we analyse a sample of 56 Type-1 AGN observed with \xmm{} and \swift{} over two decades. Observations from \xmm{} provide information on the short-time variability down to hundreds of seconds while the \swift{} monitoring allows the longer term variations to be constrained. Using \texttt{PIORAN}, we obtain estimates of the power spectra of these sources over a broad range of frequencies. We are particularly interested in estimating the power spectral bend frequency and comparing this to other source properties. 

\section{Observations and data reduction}
\label{sec:obs}
\subsection{The sample}

In this work, we select 56 AGN categorised as Seyfert 1 galaxies having \xmm{} and \swift{} observations, and known to show strong X-ray variability based on previous works. The \xmm{} data sample timescales of few hundreds of seconds to about day, and observations separated by weeks or years probe longer timescales. The \swift{} data are necessary to cover timescales of days to year. Out of the 56 sources of our sample, 34 are present in the sample of \cite{2012A&A...544A..80G}.

The sources are presented in \cref{tab:long}. The sample collected here is not representative of the total AGN population as it contains only some of the most variable AGN and luminous in X-rays. Except for RX~J0439.6-5311, most sources are nearby objects with a redshift $z\lesssim 0.1$.

We collected black hole masses and optical luminosities at $5100$\,\AA{} from the literature. Note that we were unable to confirm that host galaxy emission had been removed correctly in all cases. Black hole masses estimates are available, based on a range of methods. The most commonly used method is reverberation mapping, but even here there are differences in assumptions and details between different papers. We decide to re-calibrate the black holes masses using optical reverberation mapping measurements when available or scaling relation otherwise. We apply consistent calibration on all sources and measurements to obtain consistent estimates, this is detailed in \cref{apdx:calibmasses}. NGC~5506 is the only source where we could not find a reliable luminosity measurement so its black hole mass was estimated using a $M-\sigma$ relation as detailed in \cref{apdx:calibmasses}. The calibrated masses and continuum luminosities are presented in \cref{tab:long}. 

In this work, we focus on variability in the soft energy band $0.3-1.5$\,keV from \xmm{} and \swift{}, as this band is less prone to contamination from background flares compared to the hard band and shows a high signal-to-noise ratio for most of the sources in the sample. This is particularly important for the \swift{} data; the lower count-rate - compared to the \xmm{} EPIC - makes those data more sensitive to background. However, this band can be sensitive to changes in the obscuration, which should be limited in our sample of Seyfert~1 galaxies. This is this first study dedicated to the soft X-ray variability as most of the work on long-term AGN variability is based on \textit{RXTE} data, sensitive over $2-10$\,keV \citep[e.g.][]{2002MNRAS.332..231U,2003ApJ...593...96M}.

\subsection{\textit{XMM--Newton}}

We reduce data from the EPIC-pn camera \citep{2001A&A...365L..18S} using \texttt{SAS 20.0} \citep{2004ASPC..314..759G}. For consistency, all sources are extracted from circular regions of 25 arcseconds, and the backgrounds are extracted in source-free circular regions of 50 arcseconds on the same chip. Single and double events are collected in event lists using \texttt{PATTERN<=4}. Light curves in the $0.3-1.5$\,keV band are produced by accumulating events in bins of duration $\Delta t =150$\,s. Corrections for bad time intervals and loss of exposure are applied as presented in \citetalias{2025MNRAS.539.1775L}. As some observations could have been made using different modes, the final light curves are rescaled to account for variation of the live time between observations. This is done by adjusting the count-rate for the known live time of the mode. We do not correct for variations of the filter between observations as these corrections depend on the spectral shape and will be different for different objects and at different times. Observations with suspicion of pile up are discarded from the analysis. Details about the data collected, such as observation identifiers, net exposure, number of points, are available online in Table 4 of the VizieR catalogue.

For each \xmm{} EPIC-pn observation, we also extract source and background spectra in the $0.2-12$\,keV band. The spectra are grouped to have a minimum signal-to-noise ratio of five and ensuring that the width of the bins is not narrower than one-third of the energy resolution. We use these spectra to compute unabsorbed fluxes and then the $2-6$\,keV luminosity as described in \cref{apdx:X-rayfluxes}.

\subsection{\textit{Swift}}

\label{sec:calib}
We collect \swift{} light curves from the \textit{X-ray telescope} \citep{2005SSRv..120..165B} using the online service\footnote{\url{https://www.swift.ac.uk/user_objects/}} presented in \cite{2007A&A...469..379E,2009MNRAS.397.1177E}. Except for Mkn~279, all light curves were extracted in January 2024. The light curves are binned by snapshots (spacecraft orbit), corrected for pile-up, bad time columns on the CCD and vignetting. We only use datapoints obtained in Photon Counting (PC) mode and reject datapoints with a signal-to-noise ratio (count-rate divided by error) less than 3. This filtering removes between 1 and 9 datapoints for 24 sources in our sample, with two datapoints removed on average, up to 9 datapoints for NGC~4051. Details about the \swift{} light curves are available online in Table 5 of the VizieR catalogue.

To cross-calibrate the \swift{} data with the \xmm{} data we proceed in the same way as in \citetalias{2025MNRAS.539.1775L}. If the two instruments have simultaneous observations, a scaling factor is computed as the ratio between the \swift{} count-rate value and the average of the two neighbouring \xmm{} values. When there is no simultaneous observations, we use the empirical value of $13.3$ to rescale the \swift{} data. As in \citetalias{2025MNRAS.539.1775L}, we will include a multiplicative parameter $\gamma$ in the modelling to rectify if necessary the calibration during the inference. This allows us to check for systematic issues in the cross-calibration. In practice, we find that this parameter has no noticeable impact on the power spectrum parameters.

\section{Method}
\label{sec:method}
Since the final light curves obtained from the combination of \xmm{} and \swift{} data contain gaps and can have irregular time sampling, we cannot use Fourier-based methods to estimate the power spectrum of the fluctuations. Instead, we use a time-domain method based on Gaussian process regression and which importantly is immune to irregular sampling, accounts for the error bars and has a well-defined likelihood. In the next section, we briefly summarise the method presented in \citetalias{2025MNRAS.539.1775L} and how it is applied here.

\subsection{\texttt{PIORAN}}

We assume the power spectrum can be well described by a smooth, piece-wise power law function. Additionally, we assume the function is decreasing in power density to higher frequencies and has $1-3$ distinct power law indices ($\alpha_3 \ge \alpha_2 \ge \alpha_1$) joined by smooth bends. This means some more complicated power spectral shapes including QPOs are not included in our model. Following on from the work of  \cite{2004MNRAS.348..783M} and \cite{2007MNRAS.378..649S}, we use the following function as our power spectrum model:

\begin{equation}
    \mathcal{P}(f) = N\left(\frac{f}{f_{b,1}}\right)^{-\alpha_1}\prod_{i=1}^{n} \left[{1+(f/f_{b,i})^{(\alpha_{i+1}-\alpha_{i})}}\right]^{-1}.
    \label{eq:modelPL}
\end{equation}

For $n=2$, this model has parameters for three indices ($\alpha_{i}$), two bend frequencies ($f_{b,i}$) and an overall normalisation ($N$). We approximate this function as the sum of a set of basis functions. We can then approximate the autocovariance function corresponding to our power spectral model as the sum of the autocovariance functions of the individual basis functions. This is sufficient to describe a zero-mean Gaussian process with our chosen power spectrum. 

We use $J=30$ basis functions, $\psi_j(f)$, geometrically spaced in frequency ($f_j$) from above the highest frequency accessible to our observations to below the lowest frequency directly accessible. The basis functions all have a power law index of zero below their characteristic frequency ($f_j$) and slope six (very steep) above. The approximation built from these basis functions gives a flat power spectrum to very low frequencies and a very steep power spectrum in the high frequency limit. The total power in the approximated model is always finite in our model and the power density is always positive. 

With the autocovariance function in place, we use the Gaussian process likelihood, allowing for the mean to be a non-zero constant ($\mu$). We apply this likelihood to the data that have been logarithmically transformed, and combine with a prior density on the model parameters to evaluate the model posterior. Full details of the method are given in our previous paper \citetalias{2025MNRAS.539.1775L}.

To confirm the absence of bends, we also use the unbending power law $f^{-\alpha}$ where $0<\alpha<4$. The use of this model implies that the integral of the power spectrum in the modelling can be very large, and this makes the inference very slow. In \cref{sec:varianceapdx}, we discuss an alternate parametrisation for the amplitude of the power spectrum to solve this problem. In this work, we use the fast Julia implementation \texttt{PIORAN.jl}\footnote{\url{https://github.com/mlefkir/Pioran.jl}}.

\subsection{Modelling}
\label{sec:modelling}

As discussed by \citet{2001MNRAS.323L..26U} and \cite{2005MNRAS.359..345U}, X-ray light curves from AGN and X-ray binaries often show a linear rms-flux relationship and have a seemingly log-normal flux distribution. We apply a logarithmic transformation to the data, which gives the data a more symmetric and Gaussian distribution, before using \texttt{PIORAN}. We transform the observational errors. We set the offset of the log-transformation to zero as it was found to be uninformed by the data and had no impact on the inference in early trials. We include a cross-calibration factor $\gamma$, as a free parameter between \xmm{} and \swift{} to ensure that the calibration performed in \cref{sec:calib} is accurate. As in \citetalias{2025MNRAS.539.1775L} and \citet{2016MNRAS.461.3145V} we also assess the quality of the error bars using a scale factor $\nu$. If $\nu$ is $1$, the errors are properly defined, for $\nu>1$ they are underestimated and overestimated otherwise.

We use the same prior distributions as in \citetalias{2025MNRAS.539.1775L} for the parameters of the models and follow the prior checks outlined in \citetalias{2025MNRAS.539.1775L} to ensure that the prior distributions and the approximation are suitable. As $\alpha_1$ might be steeper than $1.25$ we modify its prior to be $\mathrm{Uniform}[0,1.5]$. The prior distribution of each parameter is shown in \cref{tab:results}.

\subsection{Sampling and inference}
To estimate the posterior distribution of the parameters and obtain the value of the Bayesian evidence $Z$ we use nested sampling \citep{2004AIPC..735..395S} via \texttt{UltraNest}\footnote{\url{https://johannesbuchner.github.io/UltraNest/}} \citep{2021arXiv210109675B} with the MLFriends algorithm \citep{2019PASP..131j8005B,2014arXiv1407.5459B}. We call \texttt{UltraNest} in Julia using \texttt{PyCall.jl}, to speed-up the inference we use MPI through the Julia interface \texttt{MPI.jl} \citep{Byrne2021}. We use 400 live points to sample the parameter space. Convergence is achieved when the effective sample size is greater than 400 and the contribution of the new live points to the evidence is less than $1\%$. The Bayesian evidence is used to compare two models using the Bayes factor\footnote{It should be noted that in practice one uses the Bayes factor to compute posterior odds which incorporate a prior probability on the models. The Bayes factor without prior odds is still informative as indicates the direction in which we should we update our odds in light of the data.} $\mathcal{B}$ defined as the ratio of the evidence of the two models $\mathcal{B}=Z_1/Z_2$. If $\mathcal{B}\gtrsim50$, model\,1 can be favoured over model\,2 with very strong evidence  \citep{1939thpr.book.....J,2017pbi..book.....B}; when $\mathcal{B} \lesssim 1/50$ model\,2 can be favoured over model\,1.

\subsection{Workflow}
We follow the Bayesian workflow introduced in \citetalias{2025MNRAS.539.1775L} to obtain posterior samples and compare models. Here, we summarise the workflow that is applied for each source in the sample:

\begin{enumerate}
    
    \item First, we start with a single-bending power law model, we use $J=30$ basis functions $\psi_6$ geometrically spaced in frequency between $f_0=1/T / 20$ and $f_M=10/\mathrm{min}(\Delta t)$ where $T$ corresponds to the duration of the light curve and $\Delta t$ is the minimum time separation between two points in the light curve.
    \item We take the logarithm of the light curve to make the data Gaussian. 
    \item We use the priors presented in \cref{tab:results} and check that the power spectrum model is well-approximated by the basis functions (see section 3.2.1 in \citetalias{2025MNRAS.539.1775L}).
    \item Using \texttt{UltraNest} we sample the posterior distribution and obtain an estimate of $\ln{Z}$. We check for convergence and assess the quality of the posterior samples using posterior diagnostics. We plot the posterior predictive power spectrum and the posterior distributions.
    \item If the posterior predictive diagnostics do not show evidence of a bend or if we are unsure about the bend frequency, we go back to (i) but using the unbending power law model. Once this is done, we compute the Bayes factor $\mathcal{B}$ between the power law model and the bending power law. If the bending model is favoured ($\mathcal{B} \gtrsim 50$), we can continue. As we are interested good detections of bend frequencies, if there is no strong evidence for the bending model (over the simple power law model) we stop. In the following section, we discuss possible reasons for the non-detections.
    \item We follow steps (i) to (iv) but using a double-bending model as applied to Ark~564 in \citetalias{2025MNRAS.539.1775L}. We compute the Bayes factor between the single-bending and double-bending power law model. When $\mathcal{B} \gtrsim 50$, we favour the double-bending model and check that the posterior power spectrum shows two clear bends.
\end{enumerate}

This workflow is applied to the 56 sources in the sample. The posterior power spectra and light curves are available as supplementary material here\footnote{\url{https://mlefkir.github.io/X-ray-variability-AGN/}}. 

\section{Results}
\label{sec:results}
Out of the 56 sources we analysed in the $0.3-1.5$ keV energy band, we find that for five sources the simple power law model was preferred (no bend), for 44 sources a single bend was preferred, and for seven sources the double-bending power law model was preferred. Of the latter group, three have already had claims of a second bend published based on X-ray data, and the other four are new. These are 1H~1934-063, Mkn~279 and Mkn~766 and Mkn~1044.

The five sources that show no strong evidence for a bend are IRAS~05078+1626, IRAS~09149-6206, Mkn~6, NGC~5506 and NGC~7213. These are not considered in the following analysis, as our analysis is focused on bend timescales and high-frequency slopes. Sources with absence of detection of a bend tend to have fewer observations and/or lower signal-to-noise in the soft energy band. NGC~7213 and IRAS~05078+1626 show very flat \xmm{} light curves and less than 20 datapoints from \swift{}, for these two sources, the absence of detection is very likely due to the lack of data. For IRAS~09149-6206, NGC~5506 and Mkn~6 which host $\sim10^8\,M_\odot$ black holes, the number of datapoints is comparable to sources with detections but light curves in the soft band for these three sources have the lowest signal-to-noise ratio (count-rate/error) of the sample. Mkn~79 is also not included in this analysis as the bend frequency posterior is very similar to the log-Uniform prior making it uninformed by the data.

We ran nested sampling on the double-bending power law model for all the sources of our sample but the sampler did not reach convergence for four sources after 200 million likelihood evaluations: NGC~4051, NGC~5548, NGC~4593 and RE~J1034+396. We therefore cannot make inference on the double-bending power law model of these sources. The difficulty to sample from the posterior distribution of these sources can be due to the model being incorrect, it could be that the long-term power spectrum is more complicated than a double-bending power law. Non-stationary variability could explain why the model struggled to fit, however these sources are not known for exhibiting strong non-stationary variability. Except for NGC~4593, these sources also have a large ($\gtrsim 5000$) number of datapoints which can slow down the likelihood calculation and thus the exploration of the parameter space. In practice, when the power spectrum does not have a second bend, the double-bending model reduces to the single-bend model, when $f_1=f_2$ and/or $\alpha_i = \alpha_j$.

The median values of the parameters for the single and double bending models are presented in \cref{tab:results}, where sources with strong evidence for a second bend are presented in the first seven rows of the table.

\begin{figure*}
    \centering
    \includegraphics[width=\textwidth]{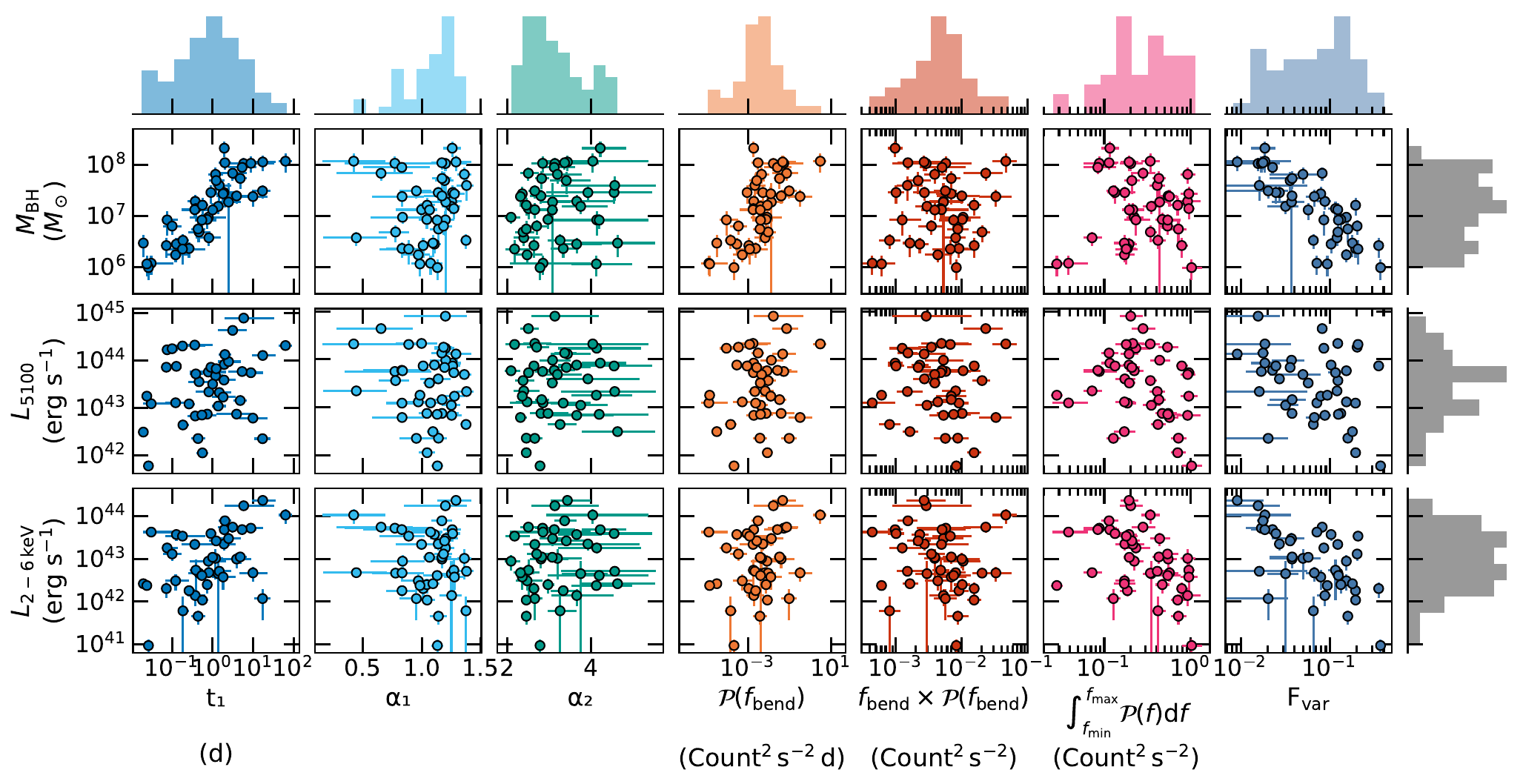}
    \caption{Correlations between the black hole mass, optical luminosity, $2-6$\,keV X-ray luminosity and the median value of the bend timescale $t_1$, low-frequency slope $\alpha_1$, high-frequency slope $\alpha_2$, power spectrum at the bend frequency, power spectrum at the bend frequency times bend frequency, integral of power spectrum model and average rms amplitude of variability $F_\mathrm{var}$ computed over segments of $30$\,ks from \xmm{} light curves. The distribution of each quantity are plotted in the upper and right panels.}
    \label{fig:correlations}
\end{figure*}

\cref{fig:correlations} shows the distribution of the parameters of the single-bending power law model and correlations against black hole mass, optical luminosity and X-ray luminosity. We only plot sources where the single-bending power law model is favoured. We recover the well-know correlation between the bend timescale and the black hole mass (top left corner) \citep{2004MNRAS.348..783M,2006Natur.444..730M}.

For each \xmm{} light curve, we compute $F_\mathrm{var}$, the square root of the excess variance \citep{2003MNRAS.345.1271V} on segments of $30$\,ks duration and calculate the unweighted average over all segments. $F_\mathrm{var}$ relates to the integrated power within the observed frequency range. We plot the averaged $F_\mathrm{var}$ in the rightmost panels of \cref{fig:correlations}. We recover the expected correlation between the amplitude of variability and the black hole mass \citep{2004MNRAS.348..207P}. 

We also compute the amplitude of the fitted power spectrum at the bend frequency and the integral of the power spectrum model between the minimum and maximum frequencies, where $f_\mathrm{min}$ and $f_\mathrm{max}$ depend on the duration of the light curve and minimum time sampling. Under the parametrisation of the approximation, the amplitude of the power spectrum is related to integral of the power spectrum from $0$ to $+\infty$. If the underlying power spectral shape is more complex than the bending power law model this parameter will poorly relate to the `true' amplitude of the power spectrum. The same goes for the integral of $\mathcal{P}$ in the observed frequency window, if the underlying power spectrum shows bumps or features not captured by the bending power law, the integral will differ from the true integrated power. We find that the power spectrum amplitude is correlated with black hole mass, \cref{apdx:psdamplitude} details how the power spectrum amplitude is computed. There are several reports that the quantity $f_b\times \mathcal{P}(f_b)$ is invariant with black hole mass; here we find this is not correlated with mass, but does show a high variance, with a mean of $0.009$ in units\footnote{The GP method used in this work models the power spectrum values in absolute units (rms$^2$\,d).} of $\mathrm{Count}^2\,\text{s}^{-2}$. As we take the logarithm of the data and model the variance in absolute units (of the log data), this variance is related to the fractional variance of the data itself\footnote{Given a positive valued time series $x$ with small fluctuations around a mean $\mu$, a first order Taylor expansion around the mean for $\ln x$ gives $\ln x\approx\ln\mu+(x-\mu)/\mu$. The variance of $\ln x$ can then be expressed as $\mathrm{Var}(\ln x)\approx \mathrm{Var }(x)/\mu^2$ which corresponds to the fractional variance.}. With this argument, this value can be compared to previous periodogram or excess variance studies \citep[e.g][]{2004MNRAS.348..207P,2012A&A...544A..80G,2023A&A...673A..68P} and agrees within error bars.

Additionally, $F_\mathrm{var}$ is computed from $30$-ks segments while the integral is calculated over the observed frequency window, which extends to lower frequencies. In \cref{apdx:integrate30ks}, we show the correlations between the integral of the power spectrum model and the black hole mass or luminosity when fixing $f_\mathrm{min}=1/{30\,\mathrm{ks}}$. We recover a dependence on mass for the integral, similar to the one with $F_\mathrm{var}$. This confirms that $F_\mathrm{var}$ and $\int_{f_\mathrm{min}}^{f_\mathrm{max}}\mathcal{P}(f)\mathrm{d}f$ convey the same information but need to be computed on a similar frequency range to be compared. This effect is similar to the one observed when computing excess variance with light curves of different duration as discussed in \cite{2023A&A...673A..68P}.

As expected from previous works, the integral of the power spectrum and $F_\mathrm{var}$ anti-correlates with the X-ray luminosity \citep[e.g][]{1993ApJ...414L..85L,2004ApJ...617..939M}. To check if there is a correlation between $F_\mathrm{var}$ and optical luminosity, we used the Spearman correlation coefficient and found no significant correlation between the two quantities. It is important to note that this sample of 43 sources with single-bend detections is small and not complete in a statistical sense, finding and interpreting correlations between parameters in this sample can therefore be challenging.

For sources with a single-bend, we find that the low-frequency slope is consistent on average with $\alpha_1\sim 1$. The high-frequency slopes, $\alpha_2$, span $2-5$ with a mean value of $\alpha_2=3$. We find that $\nu$, the scaling factor on the error bars, is close to one for most sources, except for NGC~4051, RE~J1034+396 and Ark~564. 

\subsection{Sources with evidence for two bends}

\begin{figure}
    \centering
    \includegraphics[width=\columnwidth]{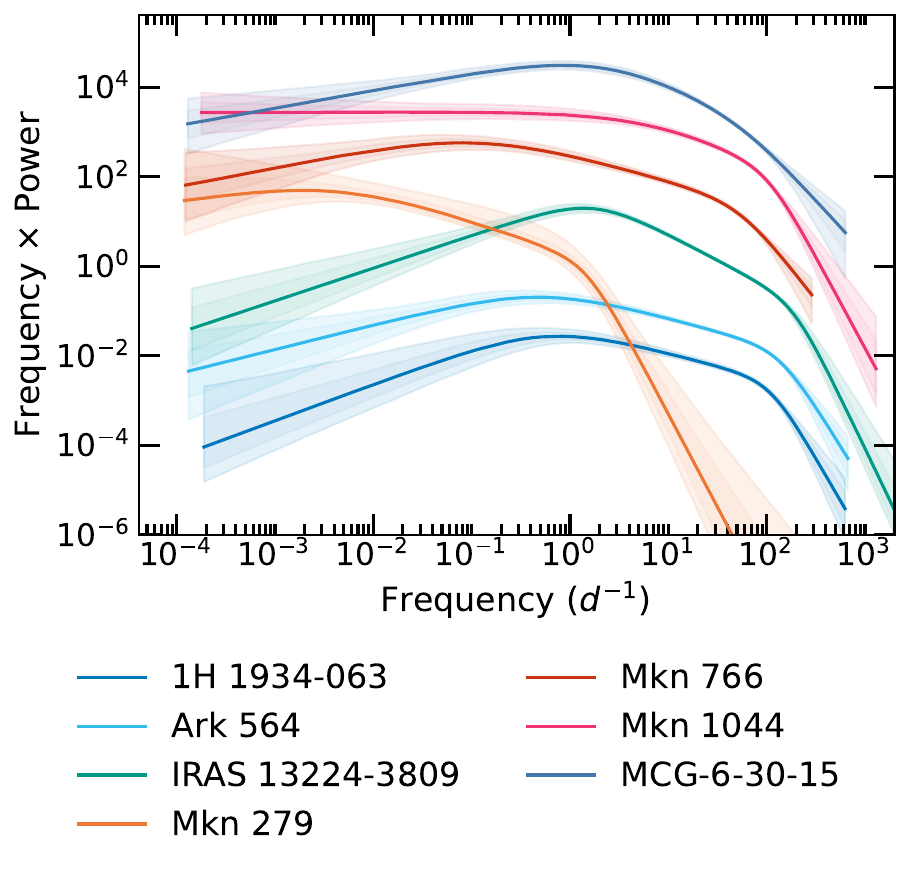}
    \caption{Posterior power spectra of the sources with evidence for two bends. The shaded areas correspond to the $68\%$ and $95\%$ percentiles from the posterior distributions. As we are mainly interested in the general shape and not total power, the amplitudes of individual sources have been rescaled for plotting purposes.}
    \label{fig:PSD_doublesources}
\end{figure}

\begin{figure*}
    \centering
    \includegraphics[width=\textwidth]{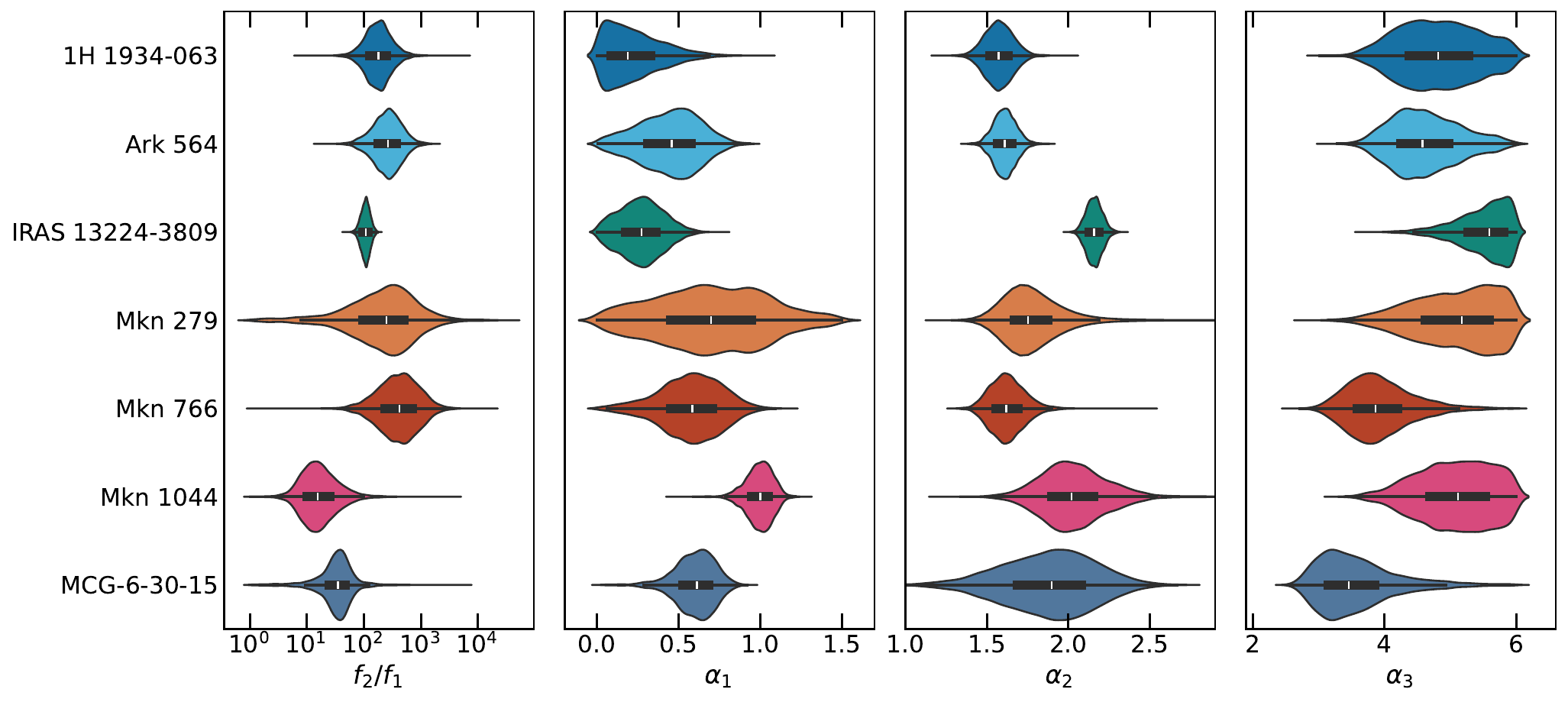}
    \caption{Distribution of the power spectral parameters for sources with evidence of a second bend. $f_2/f_1$ is the ratio between the high- and low-frequency bends and $\alpha_i$ for $i\in[1,2,3]$ are the slopes of the bending power law model.}
    \label{fig:doubleviolin}
\end{figure*}

\cref{fig:PSD_doublesources} shows the posterior model for each of the seven sources where a double-bending model is preferred. When looking at the posterior power spectrum of Mkn~1044 and MCG-6-30-15, we find that the power spectral shape is closer to a `bump' rather than two distinct bends. This is due to the separation of the two bends being an order of magnitude less than the other sources. This can be observed in \cref{fig:doubleviolin} where we plot the distribution of the ratio of the two bend frequencies for sources with evidence for a second bend. There could be several explanations for this, either the double-bending power law model is incorrect for these sources or the power spectrum is weakly non-stationary and the bends or slopes moved over time, in this situation the power spectrum we estimated is an average power spectrum. 

From \cref{fig:doubleviolin}, we observe that the low-frequency slope is less than 1 for most sources such that the power converges at low frequency without the need for any further flattening at lower frequencies. The intermediate slope is between 1 and 2.25 in most cases, and the high-frequency power spectrum is always steeper than $f^{-3}$. We note that the high-frequency power spectrum of sources with two bends is similar to the power spectrum of single-bending sources. The ratio $f_1 / f_2$ spans $100-500$.

For Mkn~279, we initially used a \swift{} light curve collected in January 2024, however given the evidence for the low-frequency bend we also included data up to April 2025 ($\sim100$ points) as a check. We find that including the recent \swift{} monitoring decreases dramatically the Bayes factor from $10^3$ to $73$, favouring less the double-bending model. This may indicate that constraints on the low-frequency bend of this source are still highly uncertain or that the power spectrum varied between January 2024 and April 2025. Further monitoring of this source may be necessary to properly constrain its power spectrum at low frequency. We note however, that the median power spectral parameters did not vary substantially with the addition of new data.

\subsection{X-ray variability plane}

\begin{figure}
    \centering
    \includegraphics[width=\columnwidth]{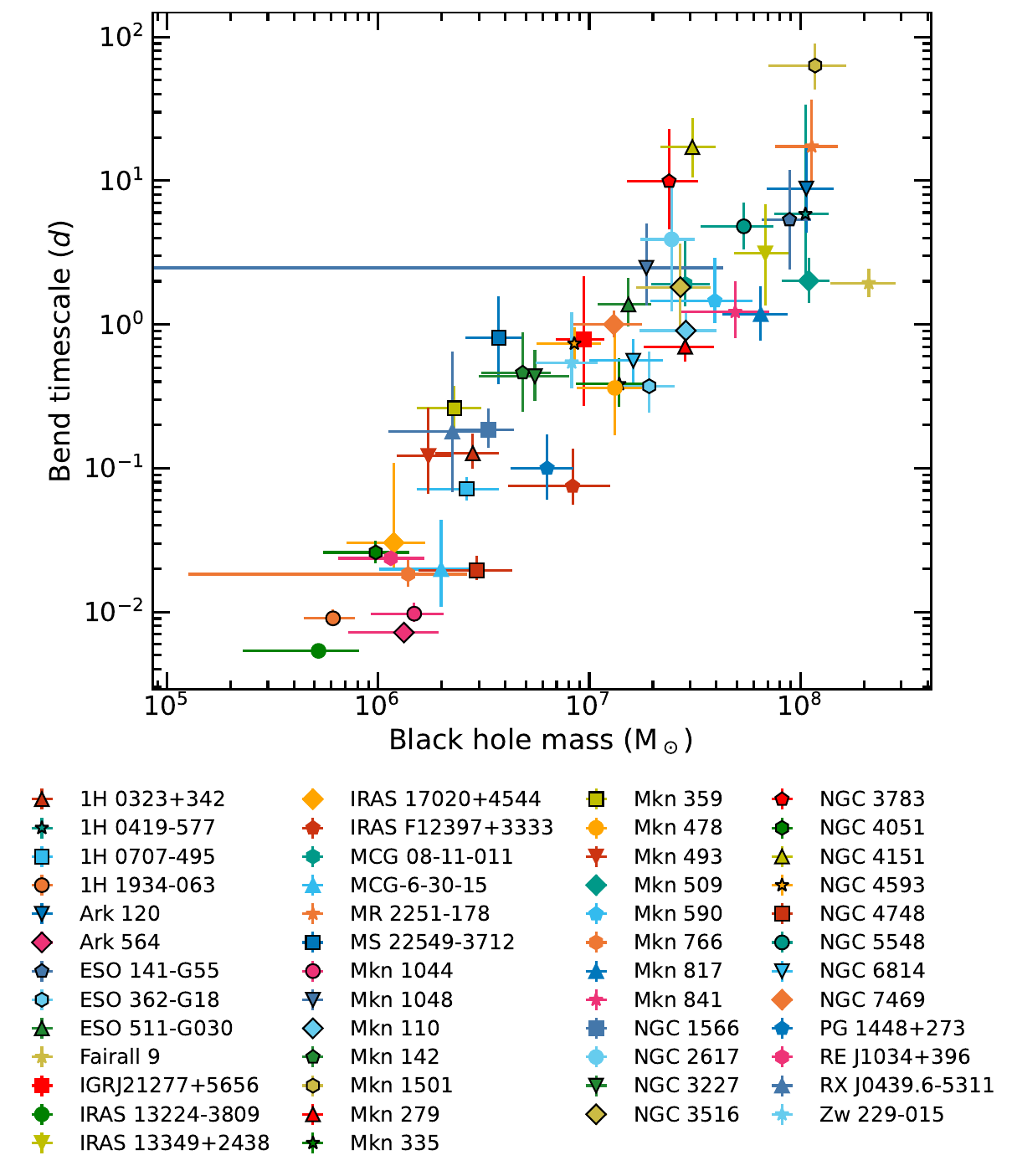}
    \caption{X-ray variability plane for all sources. For sources with detection of low-frequency bend, we use the high-frequency bend.}
        \label{fig:timescales}    
\end{figure}

We now study the correlation between the bend timescale, black hole mass and luminosity. For the 43 sources, we use $f_b$ from the single-bend model, and for the seven sources where a second bend is preferred, we use the higher frequency bend, $f_2$. We create two datasets: one with all sources and one containing only the ones with a single-bend. The correlation between black hole mass and bend timescale for the full dataset is plotted in \cref{fig:timescales}.

Following \cite{2006Natur.444..730M} and \cite{2012A&A...544A..80G}, we model the X-ray variability plane as follows:
\begin{equation}
    \log_{10}{t_\mathrm{bend}} = A  \log_{10}{M_\mathrm{BH}} + B \log_{10}{L} + C,
    \label{eq:mchardyrelation}
\end{equation}

where $M_\mathrm{BH}$ is in units of $10^6\,M_\odot$, $L$ can be the optical luminosity at $5100\angstrom$ or the $2-6$\,keV X-ray luminosity in units of $10^{44}\,$erg\,s$^{-1}$ and $t_\mathrm{bend}$ is the bend timescale in days. We define the observation model as follows:
\begin{equation}
    \log_{10}{t_\mathrm{bend,obs}} = \log_{10}{t_\mathrm{bend,model}}  + \epsilon_i
    \label{eq:observationmchardy}
\end{equation}

where  $t_\mathrm{bend,obs}$ is the measured bend timescale in the previous section, $t_\mathrm{bend,model}$ is the model timescale of \cref{eq:mchardyrelation} and $\epsilon_i \sim \text{Normal}(0,\sigma^2)$ is an intrinsic noise where $\sigma$ quantifies the intrinsic scatter in the relationship and is estimated during posterior sampling.

\subsubsection{Linear regression modelling}

We first begin to model \cref{eq:observationmchardy} without accounting for the uncertainties on any parameters. We only use the optical luminosity for this model, we will be using the X-ray luminosity in the next model. Our prior knowledge indicates a positive correlation between the mass and the bend timescale, so we choose a broad positive prior for $A$. Based on previous work, we choose simple distributions over real numbers for $B$ and $C$. As $\sigma$ is a standard deviation it must be positive. We use priors of the form:
\begin{align}
    A &\sim \mathrm{LogNormal}(0,1)\quad\quad    B \sim \mathrm{Normal}(0,2^2)\\
    C &\sim \mathrm{Normal}(-2,4^2) \quad\quad
    \sigma \sim \mathrm{Exponential}(1)
    \label{eq:priorssimple}
\end{align}

where we write the parameters of a normal distribution as $\text{Normal}(\mu,\sigma^2)$. We sample the posterior distribution using Hamiltonian Monte Carlo (HMC) with the No-U-Turn Sampler (NUTS) \citep{2011arXiv1111.4246H} implemented in \texttt{NumPyro} \citep{phan2019composable,bingham2019pyro}. We use HMC with NUTS as it can handle models with a large number of parameters and little tuning, which is necessary for the model of the following section. We run the sampler on 16 chains with 2000 iterations of warm up and 2000 iterations of sampling. We assess convergence by looking at the trace plots and using standard diagnostics such as $\hat{R}\simeq 1$ and ensuring that the effective sample size (ESS) is high enough \citep{Vehtari_2021}. 

\begin{figure}
    \centering
    \includegraphics[width=\columnwidth]{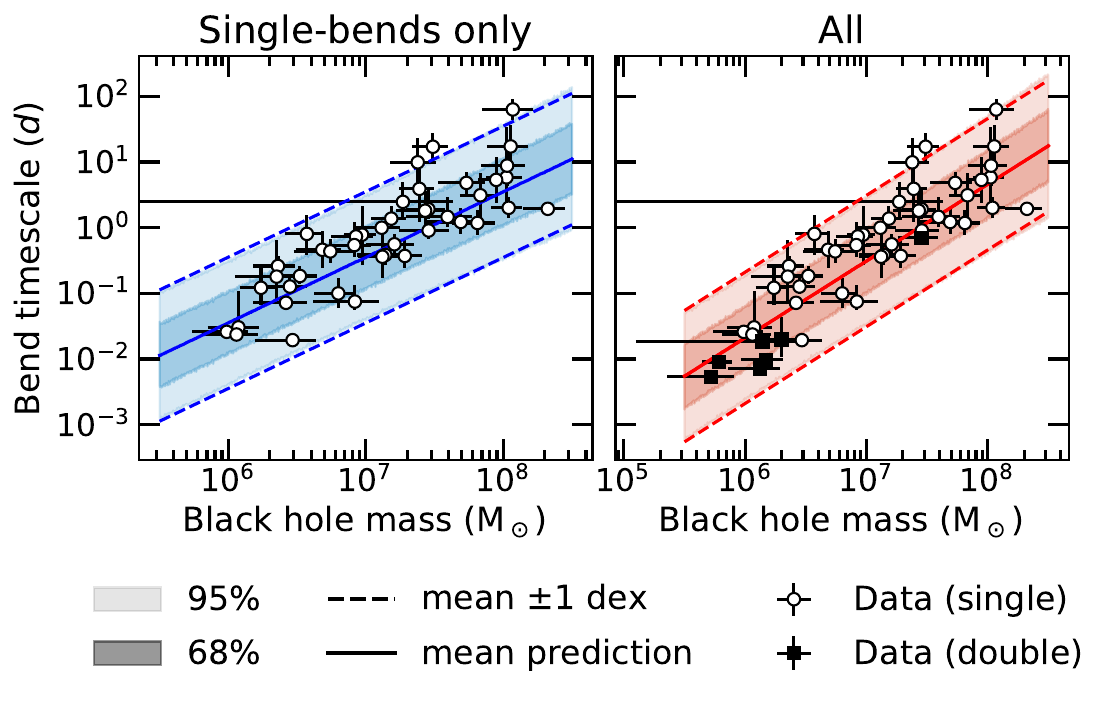}
    \caption{Prediction for \cref{eq:mchardyrelation} when using a simple modelling on the single-bend only sources (left) and all of the sources (right). The mean prediction is plotted in solid line and the $\pm 1$\,dex intervals are shown in dashed lines. The $68\%$ and $95\%$ credible intervals are shown in shaded areas. Single-bend timescales are shown in white dots and double-bend timescales are shown in black squares.}
    \label{fig:predictsimple}
\end{figure}

We plot the distributions of posterior samples in \cref{fig:contours} labelled as `simple' in dark blue and magenta. The dataset containing only sources with a single bend is mentioned with `(single)'. We find $A\simeq1.2$, $B\simeq -0.15$, $C\simeq-1.8$ which is in agreement with a previous analysis of \xmm{} data from \cite{2012A&A...544A..80G}. In \cref{fig:predictsimple}, we plot the predicted relationship for the two datasets. Except for NGC~4151, NGC~3783 and Mkn~1501 all sources fall in the $95\%$ prediction region. We observe that the two datasets give slightly different posteriors for $A$, $C$ and $\sigma$.

\subsubsection{Errors-in-variables (EIV) modelling}

\label{sec:eiv}
The linear regression modelling did not include the uncertainties on masses and bend timescales, we now use errors-in-variables (EIV) modelling\footnote{See section 15.1 of \cite{2020srbc.book.....M}.} to account for the uncertainties on the measurements. EIV modelling enables the inference of the parameters of interest of the model, here $A$, $B$, $C$ and $\sigma$, while incorporating error bars on the observed quantities $M_\mathrm{BH}$, $L$ and $t_\text{b}$. This modelling treats the mass, luminosity and timescale as latent variables, where we assume the logarithm of these quantities to be normally distributed. The diagram corresponding to the probabilistic model is shown in \cref{fig:dag}. The parameters of the model are shown in orange circles, the latent variables are represented by orange ellipses located the rectangular box. The priors on $A$, $B$, $C$ and $\sigma$ are the same as before. We assume a standard error of $0.5\,\text{dex}$ on the logarithm of the optical luminosity. For the X-ray luminosity, we use the standard error obtained from the spectral fitting. We follow the same process as before for the sampling and the convergence checks of the posterior samples. 

\begin{figure}
    \centering
    \includegraphics[width=\columnwidth]{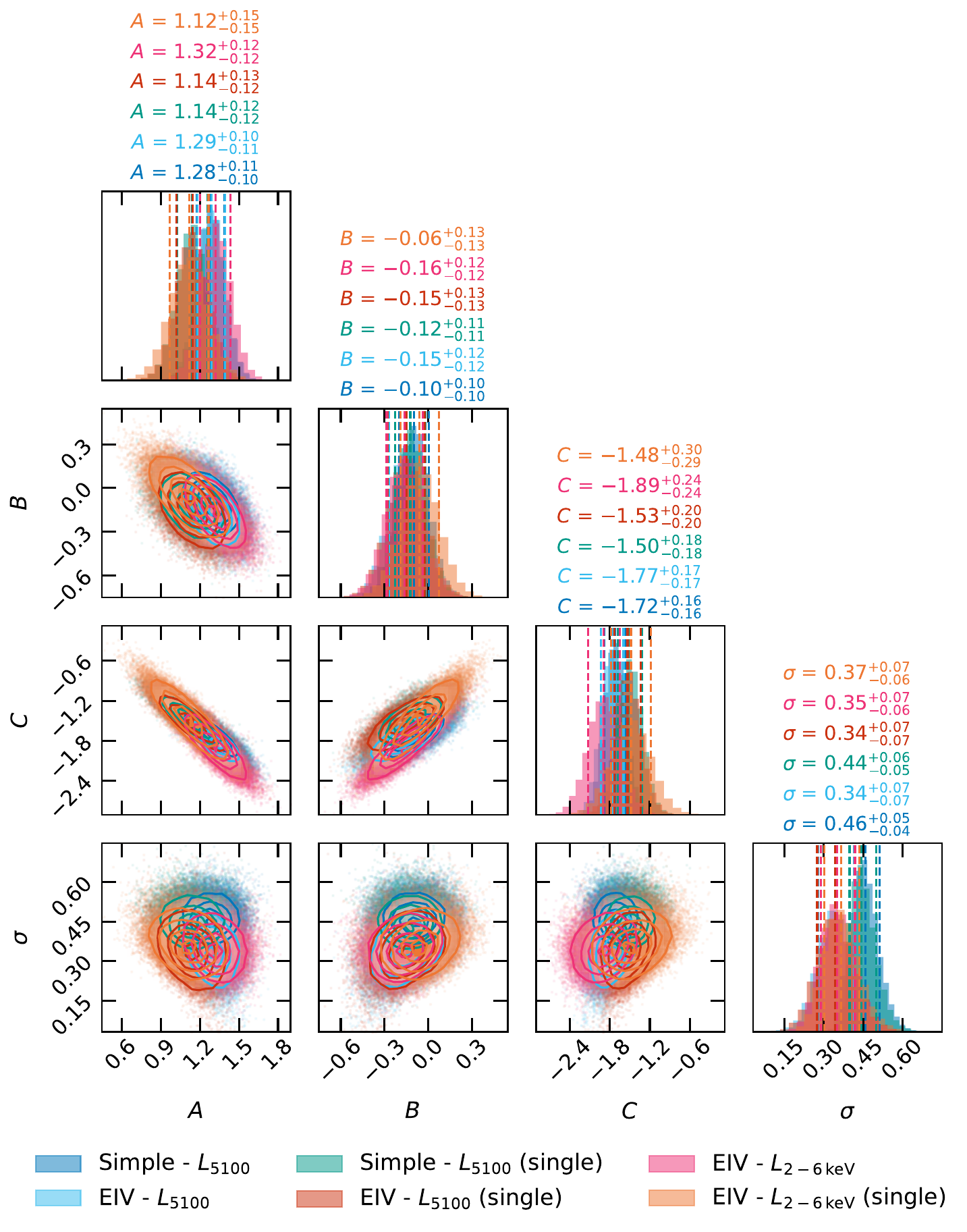}
    \caption{Posterior samples for the parameters of \cref{eq:mchardyrelation} using simple and errors-in-variables (EIV) modelling. "single" indicates that only sources with a single-bend detection are included. }
    \label{fig:contours}
\end{figure}

\cref{fig:contours} shows the distribution of the posterior samples. For each dataset, single-bend sources or all sources, the distributions for $A$ obtained with EIV modelling agree with the ones from linear regression. For $B$, the distributions are consistent across methods and datasets and agree with $B\simeq-0.15$. The intrinsic scatter $\sigma$ appears to be smaller when using the EIV modelling.

\section{Discussion}
\label{sec:discuss}

In this sample, we estimate characteristic timescales of the X-ray variability for 50 AGN by modelling a bend in the power spectrum of each. The logarithm of the bend timescale correlates linearly with the logarithm of black hole mass, however the slopes of the power spectrum do not appear to be correlated with the luminosity or mass. 

\cite{2011ApJ...730...52K} studied the X-ray variability of 10 AGN observed with \textit{RXTE} and \xmm{} using a Gaussian process method akin to the one used in this work, see the discussion of \citetalias{2025MNRAS.539.1775L} for a comparison. Our sample contains all the sources studied by \cite{2011ApJ...730...52K}. They obtained reliable bend frequency estimates for Fairall~9 and NGC~5548 for the first time, our results broadly agree with their findings within error bars. They find low-frequency slopes $\alpha_1$ between $0.7$ and $1.6$ and do not observe any correlation of the slopes with mass.

Within our sample, 34 AGN overlap with \cite{2012A&A...544A..80G}. We obtain estimates of the bend timescale for the first time in 17 sources within this overlapping subset. This study is focused on the soft X-ray band but a similar work could be reproduced in the hard band as the power spectrum is known to vary with energy \citep[e.g.][]{2001ApJ...554..710N,2004MNRAS.348..783M,2011MNRAS.413.2489V}. The sources selected for this sample are unabsorbed AGN but it should be noted that the soft band can easily be affected by changes in the absorption.

\subsection{Double bends}

Of the 56 AGN studied here, seven showed strong preference for a double-bending power law model. 

Ark~564 is a narrow-line Seyfert 1 galaxy known for the evidence of a low-frequency bend in its power spectrum \citep{2001ApJ...550L..15P,2007MNRAS.382..985M}. Ark~564 was used as a test case in \citetalias{2025MNRAS.539.1775L} where we obtained consistent results with previous Fourier-based analyses. 1H~1934-063 is a variable narrow-line Seyfert 1 which had little attention over the years except in recent works by \cite{2018ApJ...867...67F} and \cite{2022MNRAS.513.1910X}. We find that this source has a power spectrum very similar to Ark~564. 

\cite{2019MNRAS.482.2088A} studied the dramatic X-ray variability of IRAS~13224-3809 and found evidence of a low-frequency bend using standard periodogram analysis and CARMA modelling \citep{2014ApJ...788...33K}. Using the same data our results are broadly in agreement with their analysis. They found that the power spectrum changes during the monitoring campaign, suggesting non-stationarity. They also showed that the power spectrum parameters vary with energy, as already observed in various sources \citep[e.g.][]{2001ApJ...554..710N,2004MNRAS.348..783M}. Mkn~766 has been the subject of many timing studies often focused on the high-frequency power spectrum \citep[e.g.][]{2001ApJ...562L.121B,2007ApJ...656..116M,2003MNRAS.341..496V}. Here we find that the low-frequency power spectrum breaks at around $0.1\,\mathrm{d}^{-1}$.

Mkn~279 a narrow-line Seyfert 1 harbouring a $10^7 M_\odot$ black hole, may have a low-frequency bend associated with a timescale of $175^{+380}_{-140}\, \mathrm{d}$. As discussed before, the evidence for a low-frequency bend was reduced with the addition of recent \swift{} observations, therefore further monitoring will be required to properly constrain this timescale.

MCG-6-30-15 was observed by \textit{RXTE} and \textit{ASCA} and claims of a low-frequency bend around $10^{-5}\,\text{Hz}$ (to a slope $\alpha_1 \sim 0$) were made \citep{2000ApJ...531L..13N,2000MNRAS.318..857L} but disputed by \cite{2002MNRAS.332..231U}. In our analysis of Mkn~1044 and MCG-6-30-15, the Bayes factor may be high enough to consider a second bend, but we observe a `bump' in the predictive power spectrum as the two bend frequencies are close together. In agreement with \cite{2007MNRAS.378..649S} we do not find evidence for a second bend in the power spectrum of NGC~3783 using independent data.

For sources where only a single-bend provides the best explanation of the data, the bend corresponds to the high-frequency bend in the two-bend model. Where there is one bend, the high-frequency slope is greater than 2, as it is for the highest frequency slope in the two-bend models. We henceforth assume that where only one bend is required to fit the available data, that this is the high-frequency bend and it is the lower frequency bend (assuming one exists) that had not been detected.

\begin{figure}
\centering
\includegraphics[width=\columnwidth]{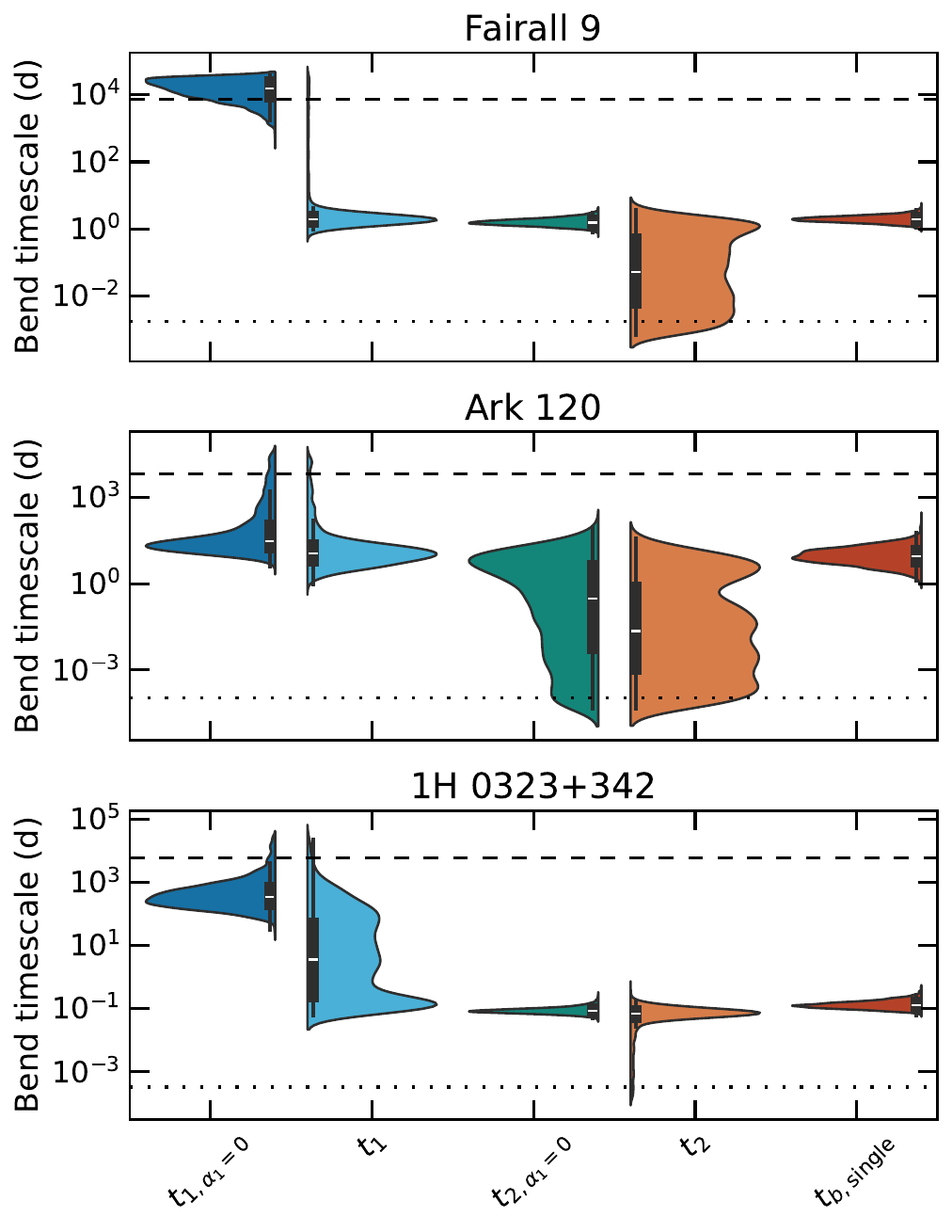}
\caption{Examples of bend timescale posteriors for the double-bend power law model when $\alpha_1$ is free and when it is set to $\alpha_1=0$ and for the single-bend model. The duration and minimum sampling of the light curves are respectively shown as dashed and dotted lines.}
\label{fig:limits}
\end{figure}

While we do not find evidence for a low-frequency bend in the other sources, we try to set lower limits on $t_1$, the low-frequency bend timescale. We run the inference on the double-bending power law model and constrain the low-frequency power spectrum to be flat, by setting $\alpha_1=0$. Using the posteriors on the bend timescales, we illustrate in \cref{fig:limits} the three types of results we find by applying this constraint. We find that for several sources, the bend is pushed to the duration of the long-term light curve. This is shown for Fairall 9 in the top panel of \cref{fig:limits}. In some cases, $t_1$ is within the observed window but overlaps with the high-frequency bend, this is illustrated in the middle panel of \cref{fig:limits}. We see that the posteriors with and without the constraint are similar. In the bottom panel, we show the case where the posterior on $t_1$ is within the observed timescales and does not overlap with the posterior on $t_2$. In this case, we can set the lower limit on $t_1$ to be the posterior mean. \cref{tab:limits} presents limits for sources in the first and third cases.

\begin{table}
\centering
\caption{Lower limits on the low-frequency bend timescale in the case of a double-bending power law model, $T$ is the duration of the light curve in days for comparison. The lower limit is computed as the posterior mean on the bend timescale in the case where $\alpha_1$ is set to $0$.}
\label{tab:limits}
\begin{tabular}{lll}\toprule
Source & Lower limit on $t_1$ (d) & $T$ (d)\\\midrule
1H~0323+342 & 889 & 6073 \\
1H~0419-577 & 2943 & 8241 \\
1H~0707-495 & 7667 & 7514 \\
ESO~362-G18 & 625 & 6287 \\
ESO~511-G030 & 13389 & 6008 \\
Fairall 9 & 16851 & 7528 \\
IGRJ21277+5656 & 149 & 6410 \\
IRAS~17020+4544 & 2267 & 6578 \\
IRAS~F12397+3333 & 1996 & 6591 \\
MCG 08-11-011 & 2983 & 6957 \\
MR 2251-178 & 6818 & 8306 \\
Mkn~1048 & 5025 & 7159 \\
Mkn~110 & 8676 & 4745 \\
Mkn~142 & 372 & 5413 \\
Mkn~335 & 13127 & 6449 \\
Mkn~359 & 1210 & 6346 \\
Mkn~478 & 872 & 7224 \\
Mkn~493 & 432 & 6870 \\
Mkn~509 & 1725 & 7868 \\
Mkn~590 & 4881 & 7920 \\
Mkn~817 & 8469 & 5968 \\
Mkn~841 & 12891 & 8248 \\
NGC~1566 & 9552 & 5804 \\
NGC~2617 & 2927 & 3706 \\
NGC~3227 & 3682 & 5684 \\
NGC~3516 & 3008 & 8185 \\
NGC~3783 & 406 & 8059 \\
NGC~4051 & 357 & 4791 \\
NGC~4151 & 7721 & 8258 \\
NGC~4593 & 10744 & 7135 \\
NGC~4748 & 6066 & 6567 \\
NGC~5548 & 14581 & 8130 \\
NGC~6814 & 8293 & 4950 \\
NGC~7469 & 7160 & 8292 \\
PG 1448+273 & 3025 & 7603 \\
RX J0439.6-5311 & 503 & 5581 \\
Zw 229-015 & 1646 & 4183 \\\bottomrule
\end{tabular}
\end{table}

\subsection{Characteristic timescales}

Finding a physical explanation or an origin for the bend frequencies in the power spectrum is challenging \citep{2003ApJ...593...96M}. The canonical \cite{1973A&A....24..337S} accretion disc model includes several characteristic timescales (dynamical, thermal and viscous), all of which will scale with mass as observed. These are based on hydrodynamic properties of the flow in an accretion disc \citep{2002apa..book.....F}, and we could try to relate them with the bend frequency. However, as already pointed out by \cite{2007MNRAS.382..985M} the dynamical timescale $t_\mathrm{dyn} \sim \sqrt{R^3 /GM}$ at a few gravitational radii is much shorter than the timescale associated with the high-frequency bend. In a thin disc, the viscous timescale is several orders of magnitude longer than the dynamical timescale. The thermal timescale is an order of magnitude longer (assuming a viscosity parameter $\alpha=0.1$) than the dynamical timescale and might agree with the bend frequency. the hydrodynamic timescales of the standard accretion disc might be expected to affect optical-UV emission directly. In this work, we study the X-ray variability and the connection between the X-ray emission process and the disc is still unclear. Thus, mapping X-ray bend timescales to simple disc timescales may not be accurate.
It is acceptable to believe that accretion discs are probably at least weakly magnetised, which is enough to generate turbulence via the magneto-rotational instability (MRI) \citep{1991ApJ...376..214B}. Turbulence could be the source of fluctuations in the mass accretion rate that propagates inwards in the disc \citep{2006MNRAS.367..801A}.

\cite{2020MNRAS.496.3808B} studied results from general relativistic magneto-hydrodynamic (GRMHD) simulations and computed power spectra of mass accretion-rate time series. The power spectra of these simulations typically show steep low-frequency power spectra ($\alpha_1 \sim 1.5$) with a break to a steeper power-law ($\alpha_2 \sim 2.4-3$), these values are in strong agreement with our findings. When the simulations are of high-cadence, the high-frequency power spectrum can be well-modelled and the recover a break frequency at around $10^3 M_\odot/M$~Hz which is within the same order of magnitude as our X-ray bends. They find the bend frequency does not relate to the viscous or the Keplerian frequencies but to the radial epicyclic frequency at $8$ gravitational radii. Progress in GRMHD simulations, with radiative processes, will be key to relating the power spectral breaks to physical properties of the inner region.

\subsection{Comparison with BHXRBs}

Compared to AGN, the variability from BHXRBs is usually more complex, often showing narrow and broad features in the power spectrum. BHXRBs power spectra are usually modelled with a mixture of Lorentzians. In this work, we used a double-bend power law to model the power spectrum which is an oversimplification compared to BHXRBs power spectra. \cite{2019MNRAS.482.2088A} used CARMA to model the power spectral shape of IRAS~13224-3809, this approach could be applied to other sources with evidence for two bends to model their power spectral shape in detail.
The double-bending shape observed here in AGN power spectra is similar to the hard/intermediate state of Cygnus~X-1. The separation between the two bends in the power spectrum is about two decades which is consistent with \cite{2007MNRAS.382..985M}. They state that this separation is typical in intermediate state BHXRBs. It should be noted that the separation and width of the Lorentzians varies between and during states \citep[e.g.][]{2005A&A...440..207B}. 

The power spectrum of Cygnus~X-1 in its soft state is very similar to the single-bending power spectrum observed in AGN \citep{2004MNRAS.348..783M}. \cite{2005MNRAS.364..208D} argued that this comparison between the soft state of Cygnus~X-1 and AGN can be debated as Cygnus~X-1 is not representative of the population of BHXRBs as it is a high-mass X-ray binary with persistent emission. Unlike most BHXRBs, Cygnus~X-1 has never been in quiescence since its discovery \citep{2024Galax..12...80J} and more importantly it accretes at a relatively high rate from a strong stellar wind, in opposition to Roche lobe overflow occurring in low-mass XRBs.  Due to a different type of accretion, the accretion disc around Cygnus~X-1 is likely smaller compared to low-mass XRBs. Similarly, AGN accretion discs are smaller for their mass relative to low-mass XRBs; as the disc temperature in AGN is lower such that hydrogen is ionised at smaller radii than in low-mass XRBs \citep{2007ASPC..373..149U,2005MNRAS.364..208D}. This general property of AGN is similar to the accretion process of Cygnus~X-1, and so we argue this could be a good reason for comparing the two. Additionally, QPOs can also be observed in the soft state \citep{2019NewAR..8501524I} which is uncommon for AGN power spectra. 

\subsection{X-ray variability plane}

Using \xmm{} and \swift{} we obtain 50 high-frequency AGN bend timescales spanning four orders of magnitude, from about 7 minutes to 62 days which is currently the largest sample of AGN power spectral bends in the soft X-ray band. 

\cite{2006Natur.444..730M} and \cite{2012A&A...544A..80G} collected bend timescales in the $2-10$\,keV energy range and used the bolometric luminosity as a proxy for the mass accretion rate in their relation. Here we used the $5100\angstrom$ optical luminosity and the X-ray luminosity as they can be easily measured, unlike the bolometric luminosity which requires modelling the spectral energy distribution or using a scaling relation. If one assumes linear empirical relations such as $L_\mathrm{bol}=9L_\mathrm{5100}$ from \cite{2000ApJ...533..631K}, then our values of $A$ and $B$ can be compared to previous studies as it will only change the offset $C$. Recent power law-like bolometric corrections such as in \cite{2019MNRAS.488.5185N} or \cite{2012MNRAS.422..478R} will change the values of $B$ and $C$. If we use the power-law bolometric correction of \cite{2019MNRAS.488.5185N}, then the bolometric luminosity is given by $L_\mathrm{bol}=k_\mathrm{bol}(5100) L_\mathrm{5100}$ where $k_\mathrm{bol}(5100)=c(L_\mathrm{5100} / 10^{42}\,\mathrm{erg}\,\mathrm{s}^{-1})^d$ with $c=40$ and $d=-0.2$. Using $L_\mathrm{bol}$ in \cref{eq:mchardyrelation} gives $B_\mathrm{Bol}=\frac{B_\mathrm{5100}}{1+d}=B_\mathrm{5100}/0.8$. Given that our values of $B_\mathrm{5100}$ are small and close to zero, the effect of these corrections on the correlation with optical luminosity should be negligible. As a check, we ran the analysis of \cref{sec:eiv} on the bolometric luminosities derived using the power-law bolometric correction and found that the value of $B$ agrees with the one of obtained using the optical luminosity, showing no strong correlation with luminosity.  For similar reasons, an X-ray bolometric correction such as the one from \cite{2004MNRAS.351..169M} will also have limited effect on the correlation in the X-ray variability plane.

Using the linear regression and EIV modelling we find that $A$ is close to unity and the correlation with (X-ray or continuum) luminosity  is weak compared to \cite{2006Natur.444..730M} which found $A\simeq 2$ and $B\simeq -1$. Our results are in better agreement with \cite{2012A&A...544A..80G} who obtained $A\simeq 1.3$ and $B\simeq -0.2$, where $B$ could be consistent with zero within $1\sigma$. \cite{2018ApJ...858....2G} included the spectral index $\Gamma$ and the column density of the absorber of the X-ray spectrum in the analysis of the variability plane. They recover a stronger dependence with bolometric luminosity ($B \simeq -0.8$) when excluding absorbed sources and a weak dependence ($B \simeq -0.2$) when including absorbed sources. Our modelling includes an intrinsic scatter $\sigma$ in the relation, to our knowledge, the fitting approaches used by previous authors assume this scatter to be zero. This forces the scatter observed in the X-ray variability plane to be explained only by the luminosity. As $\sigma$ does not tend to $0$ with our analysis, this suggests that the scatter in the relation cannot solely be explained by the luminosity. We note that even though a correlation between bend timescale and X-ray luminosity can be seen in \cref{fig:correlations}, this correlation originates from the correlation between black hole mass and X-ray luminosity. In the context of the X-ray variability, the correlation between bend timescale and black hole mass is stronger and as seen before, the X-ray luminosity is not needed to explain the scatter in the relationship in agreement with \cite{2012A&A...544A..80G}. \cite{2006Natur.444..730M} analysed $2-10$\,keV \textit{RXTE} data, and did not include an intrinsic scatter in the variability plane, this could explain the strong dependence on luminosity they obtained compared to this work.

As shown in \cref{fig:correlations}, we also recover the well-known anti-correlation between the rms variability amplitude and the black hole mass \citep{2004MNRAS.348..207P,2012A&A...542A..83P}. Similarly, the integral of the power spectrum over the same frequencies (with $f_\mathrm{min}=1/T$ set by $T=30$\,ks) is anti-correlated with the black hole mass.
We find a correlation between the black hole mass and the amplitude of the fitted power spectrum. We do not find that the amplitude of the power spectrum  - in frequency $\times$ power - at the bend is constant for all sources and masses. We find that the its mean value is close the value obtained in previous works \citep{2004MNRAS.348..207P,2012A&A...544A..80G,2023A&A...673A..68P}. The amplitude of the power spectrum can also depend on the Eddington ratio as suggested in previous work using the excess variance \citep{2008A&A...487..475P,2012A&A...542A..83P,2017MNRAS.471.4398P}. Following a different approach, \cite{2021MNRAS.508.3463G} found that the power spectrum amplitude may depend on mass and accretion rate.

Our results support the simplest scaling, the characteristic timescales are proportional to the black hole mass, itself scaling the size of the central $\sim10~r_g$ region. The timescales are not affected by luminosity or other properties. The simplest interpretation is that the high-frequency bend timescale is fixed by the geometry of the inner region and not physical properties of the region, such as temperature or density or magnetic field strength - as these will vary across our AGN sample, and be significantly different when we follow the relation down to BHXRBs. 

Double-bend sources show a high-frequency bend at systematically shorter timescales compared to single-bend sources. While this subset is small and the sources still fall within the $95\%$ credible interval, there could be an additional process unique to double-bend sources.

\subsection{QPOs}
Several sources in our sample have evidence or claims of QPOs in one or several \xmm{} observations. As discussed in \citetalias{2025MNRAS.539.1775L}, the bending power law model does not include narrow features, but the diagnostics can help identifying if features are missing in the power spectrum model. If a QPO is missing, one can use a celerite term \citep[e.g. SHO model in][]{2017AJ....154..220F} with the centroid, width and amplitude of the QPO as free parameters.

Claims of a transient QPO around $10^{-4}\,\mathrm{Hz}$ have been made in Mkn~766 \citep{2007ApJ...656..116M,2017ApJ...849....9Z}. We find a bend frequency in the same frequency range as the QPO but the diagnostics do not indicate a missing component and the bend frequency is similar to the one found in \cite{2010MNRAS.402..307V}. \cite{2015MNRAS.449..467A} detected a QPO at $1.5\times 10^{-4}\,\mathrm{Hz}$ in the $1.2-5.0\,\mathrm{keV}$ energy band of MS~22549-3712. In our analysis, we estimate the bend frequency to be around $1.4\times 10^{-5}\,\mathrm{Hz}$ in the $0.3-1.5\,\mathrm{keV}$. With our choice of energy range we may not be probing the quasi-periodic variability of this source. 

There is strong evidence for a QPO at $2.6\times 10^{-4}\,\mathrm{Hz}$ in the $1-4\,\mathrm{keV}$ power spectrum of RE~J1034+396 \citep{2014MNRAS.445L..16A,2008Natur.455..369G}. We find a bend at $4.9\times 10^{-4}\,\mathrm{Hz}$ which may be overlapping with the QPO. Note that the value of $\nu$ we find $(\nu\sim 1.3)$ is higher than expected which may hint that there is some unmodelled variance due to the error bars being underestimated. As the QPO may not be present in all observations and hardly detectable with \swift{}, an in-depth study of this source with short segments is more appropriate than the current long-term study.

\section{Conclusions}
\label{sec:conclusion}
We studied the long-term X-ray variability of 56 unabsorbed AGN using archival observations from \swift{} and \xmm{} over 20 years. Our results can be summarised as follows:
\begin{enumerate}
    \item We obtain estimates of the high-frequency bend in the power spectrum of 50 AGN. The corresponding timescales range four orders of magnitude between 7 min and 62 days.
    \item In all cases, the high-frequency power spectrum is steeper than $f^{-2}$ and the low-frequency power spectrum is steeper than white noise. This rules out damped random walk as a model for the X-ray fluctuations on all timescales. Using an alternative method, \cite{2023A&A...673A..68P} also found that the high-frequency X-ray power spectrum of local and high-redshift AGN is steeper than a damped random walk.
    \item We find three new sources with an additional bend located at low frequencies: 1H~1934-063, Mkn~766 and Mkn~279.
    \item We have refined the X-ray variability plane of \cite{2006Natur.444..730M} and find, in agreement with \cite{2012A&A...544A..80G}, that the dependence with continuum or X-ray luminosity is weak. When assuming simple or recent power-law bolometric corrections, the correlation of bend timescale with bolometric luminosity remains weak in the variability plane.
    \item We recover the correlations between black hole mass and excess variance or integral of the power spectrum (when integrated over the same frequencies), and do not find the amplitude of the power spectrum (multiplied by frequency) at the bend frequency to be constant. We find its distribution to be scattered within a factor of ten around the mean value of $0.009$ in agreement with previous works \citep{2023A&A...673A..68P,2012A&A...544A..80G,2004MNRAS.348..207P}.
\end{enumerate}

The method introduced in \citetalias{2025MNRAS.539.1775L} has been successfully applied to a small sample of unabsorbed AGN observed in the soft X-ray band. In a future paper, we will repeat this analysis in the UV/optical using a subset of this sample using data from \swift. This will allow a study of the wavelength dependence of the power spectrum. Upcoming large sky surveys, like Vera Rubin/LSST \citep{2019ApJ...873..111I} will provide a tremendous amount of data that will enable study of the long-term optical variability. 

\section*{Acknowledgements}

We thank the anonymous referee for providing comments that greatly improved this work. ML thanks S. Dupourqu{\'e} for his help on using \texttt{jaxspec}. ML acknowledges discussions on correlations with M. Ayachi.
ML acknowledges support from STFC studentships. This research has made use of the SIMBAD database, CDS, Strasbourg Astronomical Observatory, France \citep{2000A&AS..143....9W}. This research has made use of the VizieR catalogue access tool, CDS, Strasbourg Astronomical Observatory, France. This research used the ALICE High-Performance Computing Facility at the University of Leicester. This work has made use of observations obtained with \xmm{}, an ESA science mission with instruments and contributions directly funded by ESA Member States and NASA. This work also made use of observations and data supplied by UK \swift{} Science Data Centre and has made use of data and/or software provided by the High Energy Astrophysics Science Archive Research Center (HEASARC), which is a service of the Astrophysics Science Division at NASA/GSFC.

\textit{Julia Packages}: MPI.jl \citep{Byrne2021}, Julia \citep{Julia-2017}, Turing.jl \citep{ge2018turing}.

\textit{Python libraries}: arviz \citep{arviz_2019},  corner \citep{2016JOSS....1...24F}, jaxspec \citep{2024A&A...690A.317D}, matplotlib \citep{Hunter:2007}, numpy \citep{harris2020array}, numpyro \citep{phan2019composable},  ultranest \citep{2021arXiv210109675B},.

%%%%%%%%%%%%%%%%%%%%%%%%%%%%%%%%%%%%%%%%%%%%%%%%%%

\section*{Data Availability}

The data used in this paper are publicly available to access and download from the \xmm{} Science Archive and the UK \swift{} Science Data Centre. The tables and list of observations are available in a VizieR catalogue available at CDS via anonymous ftp to cdsarc.u-strasbg.fr (130.79.128.5) or via \href{https://cdsarc.cds.unistra.fr/viz-bin/cat/J/MNRAS}{https://cdsarc.cds.unistra.fr/viz-bin/cat/J/MNRAS}. Final data products from this study can be provided on reasonable request to the corresponding author.



\bibliographystyle{mnras}
\bibliography{refs} % if your bibtex file is called example.bib

%%%%%%%%%%%%%%%%%%%%%%%%%%%%%%%%%%%%%%%%%%%%%%%%%%

%%%%%%%%%%%%%%%%% APPENDICES %%%%%%%%%%%%%%%%%%%%%

\appendix

\section{The sample}

\subsection{Calibrating black hole masses}

\label{apdx:calibmasses}
Estimating masses of supermassive black holes at the centre of active galaxies is challenging and often requires good calibration \citep{2014SSRv..183..253P}. In reverberation mapping campaigns, the black hole mass is estimated using \cref{eq:MBH}, where $R_\text{BLR}$ is the size of the broad-line region (BLR) and $\Delta V$ is the velocity dispersion of a variable line emitted from the BLR. $R_\text{BLR}$ is usually measured using the time delay $\tau$ between the emission of the H$\beta$ line and a reference continuum band, usually around $5100\angstrom$. $f_\text{BLR}$ is a scaling factor which contains most of the geometry and inclination of the system that we cannot resolve spatially. 

\begin{equation}
    M_\mathrm{BH} = f_\text{BLR} \dfrac{\Delta V^2 R_\text{BLR}}{G}.
    \label{eq:MBH}
\end{equation}

In this work, we re-calibrate the black hole masses of the sample to obtain consistent estimates for all sources and methods. In our sample, 33 of 56 sources have reverberation mapping measurements and 22 sources can have their mass estimated using scaling relations. The mass of NGC~5506 is estimated using the $M-\sigma$ relation from \cite{2011Natur.480..215M} with a velocity dispersion of $175.4\,\text{km}\,\text{s}^{-1}$ obtained from the HyperLEDA\footnote{\url{http://atlas.obs-hp.fr/hyperleda}} catalogue \citep{2014A&A...570A..13M}.

\subsubsection{Reverberation mapping} 
For sources with reverberation mapping campaigns, we collect $\tau_\text{ICCF}$ the rest-frame time delays  of the H$\beta$ line with respect to the $5100\angstrom$ continuum. For internal consistency, we collect only measurements made from the centroid of the interpolated cross-correlation function (ICCF) \citep{1986ApJ...305..175G,1998PASP..110..660P}. The line dispersion can be computed on the mean $(M)$ or the rms $(R)$ spectrum using two metrics, either the line width or the full-width at half maximum (FWHM). When possible we collect $\sigma_R$ the width from the rms spectrum. For IRAS~F12397+3333, Mkn~493 and NGC~2617, we could only find $\sigma_M$, so we compute $\sigma_R$ using the method and coefficients of Table 3 from \citet{2020ApJ...903..112D}. 
Mkn~1044 and 1H~0323+342 only have measurements of $\text{FWHM}_M$ so we use the conversion $\sigma_M = \text{FWHM}_M/2.355$, with the caveat that the line may not be Gaussian, then we apply the same relation as before to obtain $\sigma_R$.

We compute $R_\text{BLR}$ as $c\tau_\text{ICCF}$ where $c$ is the speed of light. We assume a virial factor $f_\text{BLR}=4.3\pm 1$ following \cite{2013ApJ...773...90G}. We average the lower and upper uncertainties on the time delay to obtain an approximate uncertainty on the time delay. Finally, we compute masses using \cref{eq:MBH} for all measurements and take the mean value for each source. The reverberation mapping measurements used for the recalibration are available online in Table 1 of the VizieR catalogue.

\subsubsection{Scaling relation}
\label{sec:scalingrelation}
For sources lacking reverberation mapping data, we collect line widths and follow the same process as before to obtain the width of the rms spectrum. It should be noted that all measurements are given as the FWHM of the line in a single-epoch spectra so the conversion to $\sigma_R$ is analogous to Mkn~1044 and 1H~0323+342. To obtain $R_\text{BLR}$, we collect continuum luminosities at $5100\angstrom$ and we use the radius-luminosity relation derived in \cite{2013ApJ...767..149B}.  When no uncertainty on the FWHM is provided we assume an uncertainty of $150$\,km\,s$^{-1}$.

For IRAS~13224-3809 and IRAS~17020+4544 we did not find reliable values for the luminosity $\lambda L_{5100}$, instead we use the H$\beta$ luminosity to obtain an estimate of $\lambda L_{5100}$ using relation and values from the fifth row of Table 2 in \cite{2020ApJ...903..112D}. For IRAS~13224-3809, we compute $L_{\text{H}\beta}=4\times10^{41}\,$erg\,s$^{-1}$ using the flux value of \citep{1993A&A...279...53B} with the luminosity distance $D_L=298.7$~Mpc. For  IRAS~17020+4544  we obtain $L_{\text{H}\beta}=2.75\times10^{44}\,$erg\,s$^{-1}$ from \cite{2016A&A...591A..88B}. The black hole mass is estimated as before, we use the same virial factor, when there are several epochs the final mass is the average value. The line width and luminosity measurements used for the recalibration are available online in Table 2 of the VizieR catalogue.

\section{Estimating the X-ray luminosity}
\label{apdx:X-rayfluxes}
To compute the unabsorbed X-ray luminosity of the sources of this sample, we first estimate the X-ray flux using EPIC-pn spectra. We use \texttt{jaxspec}\footnote{\url{https://github.com/renecotyfanboy/jaxspec}} \citep{2024A&A...690A.317D}, to model the X-ray spectrum from $0.3$ to $6$\,keV with an absorber \texttt{TBabs} \citep{2000ApJ...542..914W} and two power-laws. One power-law is used for modelling soft excess emission and another for the hard X-ray emission. The model can be expressed as $\texttt{TBabs} \times(\texttt{Powerlaw}+\texttt{Powerlaw})$. We model the X-ray spectrum up to $6$\,keV to reduce the effect of the reflection spectrum (including the Fe K$\alpha$ line) on the results. For each source and each observation, we model the source and background spectra consistently using a Poisson likelihood. The posterior distribution is sampled using Hamiltonian Monte Carlo via the No-U-turn sampler \citep{2011arXiv1111.4246H} implemented in \texttt{numpyro}  using 1000 iterations for warm-up and 1000 iterations for sampling with 8 chains \citep{phan2019composable,bingham2019pyro}. We check convergence using the standard diagnostics such as $\hat{R}\simeq 1$ and ensuring that the effective sample size (ESS) is high enough \citep{Vehtari_2021}. For each source, we combine all the posterior distributions and collect the unabsorbed flux in the $2-6$\, keV energy band of the hard X-ray power-law. We take the average flux as the source flux and the standard-deviation as an uncertainty on the flux.  We then compute the X-ray luminosity using the redshift values from \cref{tab:long} assuming $H_0=67.4\,\mathrm{km}\,\mathrm{s}^{-1}\,\mathrm{Mpc}^{-1}$ \citep{2020A&A...641A...6P}.

\section{Variance of the process}

\subsection{New parametrisation}
\label{sec:varianceapdx}
As introduced in \citetalias{2025MNRAS.539.1775L}, the amplitude of the power spectrum is not used as a parameter, instead we use the integral of the approximated power spectrum over frequencies from 0 to $+\infty$. This integral corresponds the variance of the process which is the autocovariance function evaluated at $\tau=0$. With this approach, we make the assumption that the power spectrum model is a good model for the variations observed in the time series. This implies that the variance of the process and the sample variance of the time series should be within the same order of magnitude. When using the bending power law model with the priors of \citetalias{2025MNRAS.539.1775L}, this assumption is correct for most power law shapes and AGN light curves. In the case of an unbending power law model $\mathcal{P}(f)=f^{-\alpha}$ where $\alpha>1$, although the integral of $\mathcal{P}$  does not converge, the integral of the approximated model will converge as the basis functions have finite integral. However, the total variance will be very high due to the large power at low frequencies and will be orders of magnitude higher than the sample variance.

A solution to mitigate this issue, is to change the parametrisation of the variance parameter and only integrate the power spectrum between the key frequencies of the observation: $f_\mathrm{min}=1/T$ and $f_\mathrm{max}=1/(2\mathrm{min}(\Delta t))$. This can be done analytically for the basis functions $\psi_4$ and $\psi_6$. Following this new parametrisation, the variance parameter is more constrained and can be used with steep power law models. However, it is important to note that this integral will not yield the sample variance, even if the power spectrum is a good model for the variability as it does not incorporate the effect of sampling.

\subsection{Calculating the power spectrum amplitude }

\label{apdx:psdamplitude}
In the following, we describe how to compute the amplitude of the power spectrum with the method of \citetalias{2025MNRAS.539.1775L}. Given an input power spectrum model, $\mathcal{P}$, we will describe the steps undertaken in the algorithm to compute the final autocovariance function and power spectrum. First, the model is evaluated at the spectral points ($f_j$ for $j=0...J$) to obtain a vector of power spectrum values: $p_j=\mathcal{P}(f_j)$. For the purpose of numerical stability when solving the linear system for the approximation, the values $p_j$ are divided by $p_0 = \mathcal{P}(f_0)$, this gives $\hat{p}_j = p_j/p_0$.
The approximated power spectrum is expressed as the sum of basis functions $\psi(f)$ with amplitudes $a_j$ (to obtain the $a_j$ one must solve a system of $J$ equations and $J$ unknowns; \citetalias{2025MNRAS.539.1775L}). Then, the integral of the approximated power spectrum is computed analytically using $\tilde{I}=\int_{a}^{b} \tilde{\mathcal{P}}(f)\mathrm{d}f$ where $a=0$, $b=+\infty$ for the standard parametrisation, and $a=f_\mathrm{min}$, $b=f_\mathrm{max}$ for the parametrisation introduced in \cref{sec:varianceapdx}. The approximated autocovariance function is normalised by $\tilde{I}$ and in the modelling, we include an amplitude parameter \texttt{norm} (also called \textit{variance}) which rescales the approximated autocovariance function. The final autocovariance function used in the GP regression is given by \cref{eq:autocov}.

\begin{equation}
    \tilde{\mathcal{R}}_\mathrm{GP}(\tau) =  \dfrac{\texttt{norm}}{\mathcal{P}(f_0)\tilde{I}} \tilde{\mathcal{R}}(\tau)
    \label{eq:autocov}
\end{equation}

Using the properties of the Fourier transform, is straightforward to see that the associated power spectrum is given by \cref{eq:psd}.
\begin{equation}
    \tilde{\mathcal{P}}_\mathrm{GP}(f) =   \dfrac{\texttt{norm}}{\mathcal{P}(f_0)\tilde{I}} \tilde{\mathcal{P}}(f)
    \label{eq:psd}
\end{equation}

\subsection{Comparing $F_\mathrm{var}$ and the integral of the power spectrum}
\label{apdx:integrate30ks}

\begin{figure}
    \centering
    \includegraphics[width=\columnwidth]{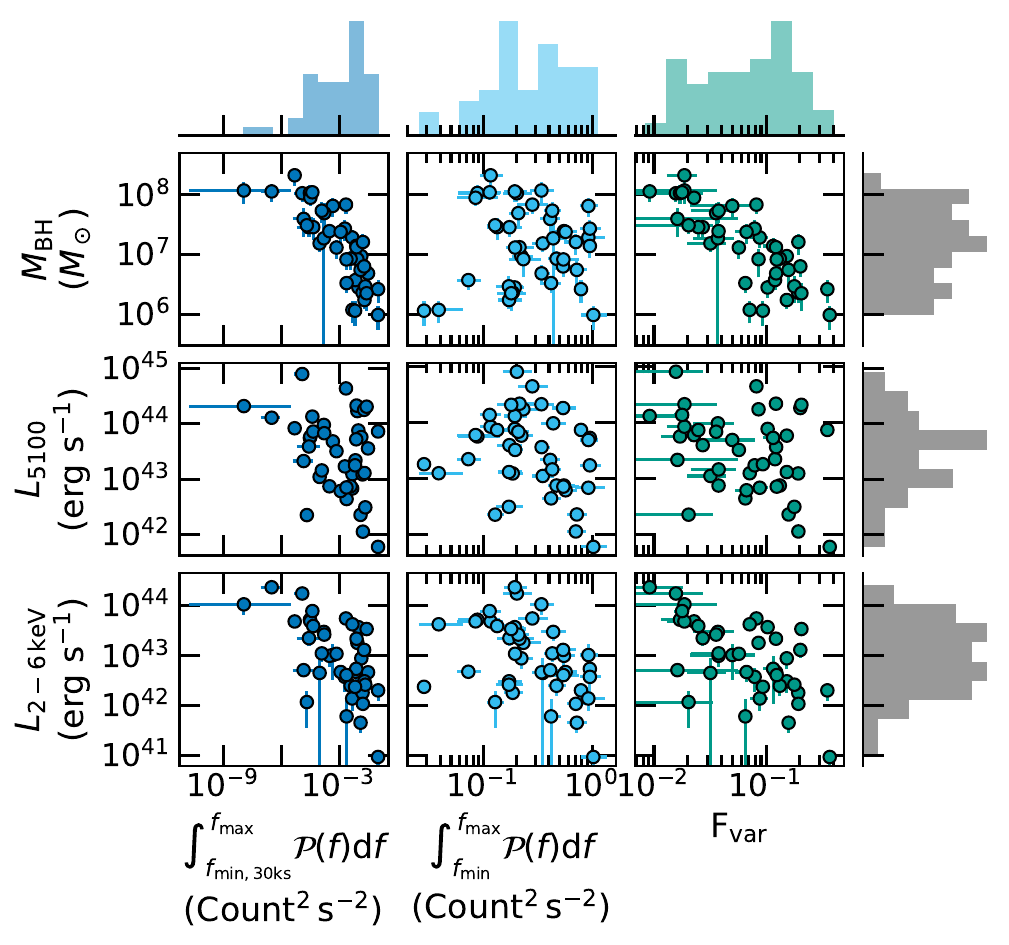}
    \caption{Same as \cref{fig:correlations} except for the first column which shows the integral of the power spectrum model between a minimum frequency defined by $1/30\,\mathrm{ks}$. }
    \label{fig:correlation30ks}
\end{figure}

In the left column of \cref{fig:correlation30ks}, we show the integral of the power spectrum but with lower bound set to $f_\mathrm{min}=1/30\,\mathrm{ks}=2.88\,\mathrm{d}^{-1}$ for all sources. It shows a clear anti-correlation with black hole mass, as with $F_\mathrm{var}$. The middle panel uses the integral over the observed frequency range and show no correlation, as already seen in \cref{fig:correlations}.

%%%%%%%%%%%%%%%%%%%%%%%%%%%%%%%%%%%%%%%%%%%%%%%%%%

\section{Tables and EIV model}
In this section, we present tables of results and details relevant to this work. We also include the EIV diagram used in \cref{sec:eiv}. \cref{tab:long} lists the physical properties of the sources analysed in this study and \cref{tab:results} lists the posterior medians with $1\sigma$ credible intervals for the parameters of the bending power law power spectra.

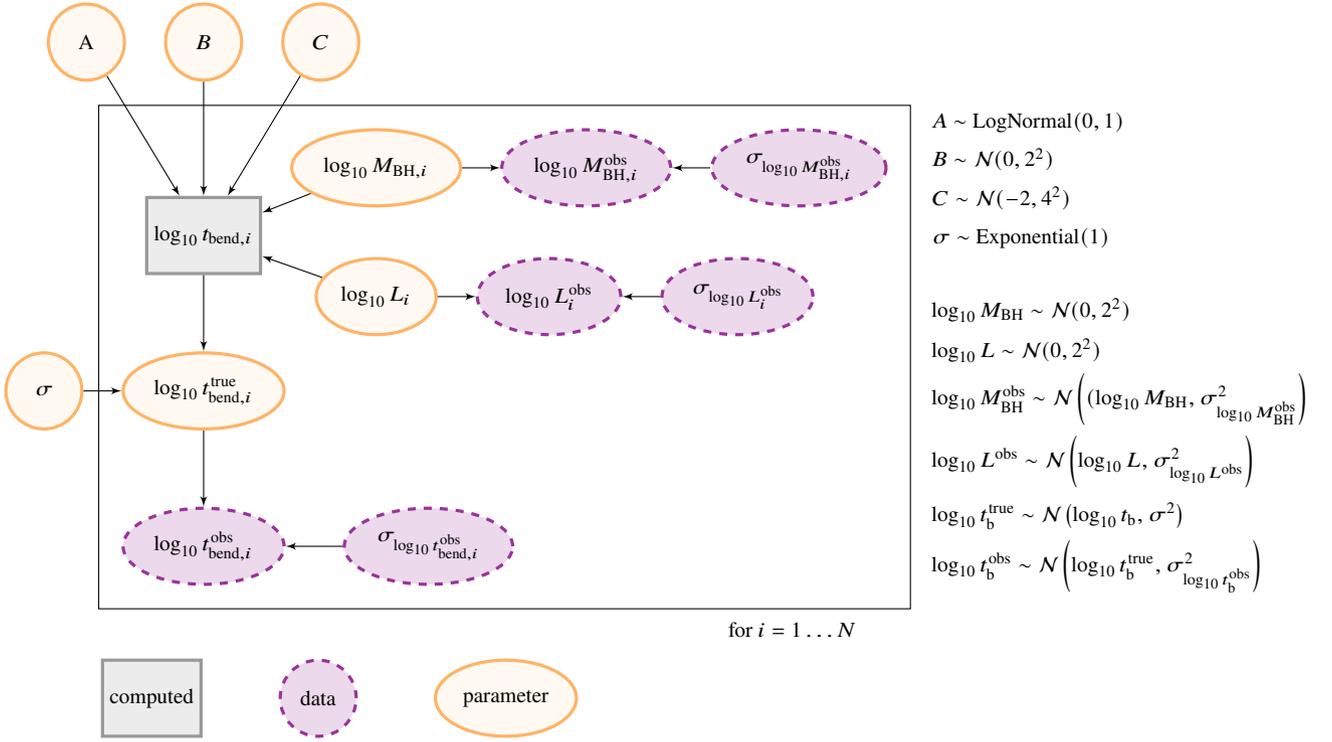
\begin{figure*}
    \centering
    \begin{tikzpicture}
        \tikzset{edge/.style = {->,> = latex'}}
        \tikzset{parameter/.style={ellipse, draw=orange!60, fill=orange!5, very thick, minimum size=1.cm,text centered}}
        \tikzset{data/.style={ellipse,dashed, draw=violet!80, fill=violet!15, very thick, minimum size=1.cm,text centered}}
        \tikzset{computed/.style={rectangle,draw=gray!80, fill=gray!15, very thick, minimum size=1.cm,text centered}}
    
        % nodes %
        \node[draw,parameter] (A) {A};
        \node[draw,parameter,right = 0.5 of A,] (B) {$B$};
        \node[draw,parameter,right=0.5 of B] (C) {$C$};
    
        \node[draw, computed, below = 1.5 of B] (Tb) {$\log_{10} t_{\mathrm{bend},i}$};
    
        \node[draw,parameter,above right = 0. and .70 of Tb] (M) {$\log_{10}M_{\mathrm{BH},i}$};
        \node[draw,parameter,below = .65 of M,] (L) {$\log_{10}L_{i}$};
    
        \node[draw,data,right = 0.5 of L,] (Lobs) {$\log_{10}L_{i}^\mathrm{obs}$};
        \node[draw,data,dashed,right = 0.5 of Lobs] (sigLobs) {$\sigma_{\log_{10}L_{i}^\mathrm{obs}}$};
    
        \node[draw,data,right = 0.5 of M,] (Mobs) {$\log_{10}M_{\mathrm{BH},i}^\mathrm{obs}$};
        \node[draw,data,right = 0.5 of Mobs] (sigMobs) {$\sigma_{\log_{10}M_{\mathrm{BH},i}^\mathrm{obs}}$};

        \node[draw, parameter, below = 1 of Tb, text centered] (Tbmod) {$\log_{10} t_{\mathrm{bend},i}^\mathrm{true}$};

        \node[draw, parameter, left = 0.5 of Tbmod, text centered] (sigTbmod) {$\sigma$};

        \draw[edge] (sigTbmod) to (Tbmod);

        \node[draw, data, below = 1 of Tbmod, text centered] (tb) {$\log_{10} t_{\mathrm{bend},i}^\mathrm{obs}$};

        \node[draw, data, right = 0.75 of tb, text centered] (sigtb) {$\sigma_{\log_{10} t_{\mathrm{bend},i}^\mathrm{obs}}$};
    
        \draw[edge] (A) to (Tb);
        \draw[edge] (C) to (Tb);
        \draw[edge] (B) to (Tb);
        \draw[edge] (sigtb) to (tb);
        \draw[edge] (Tb) to (Tbmod);
        \draw[edge] (Tbmod) to (tb);
        \draw[edge] (M) to (Tb);
        \draw[edge] (L) to (Tb);
    
        \draw[edge] (L) to (Lobs);
        \draw[edge] (sigLobs) to (Lobs);
    
        \draw[edge] (M) to (Mobs);
        \draw[edge] (sigMobs) to (Mobs);
    
    \node[draw,inner sep=3.mm,label={[label distance=1mm]-50:for $i=1\dots N$},fit=(M) (tb) (sigMobs) (sigtb)] (rectbox) {};
    
    \node[draw,parameter,below right = 0.8 and -6. of rectbox ] (param) {parameter} ;
    \node[draw,data,left=1 of param ] (dat) {data} ;
    \node[draw,computed,left=1 of dat ] (comp) {computed} ;
    
    \node[above right =-6.1 and .75 of sigMobs, text centered, align=left] (eqg) { \\\vspace{0.05cm}
$A \sim \mathrm{LogNormal}(0,1)$\vspace{0.05cm}\\\vspace{0.05cm}
$ B \sim \mathcal{N}(0,2^2) $\vspace{0.05cm}\\\vspace{0.05cm}
$C \sim \mathcal{N}(-2,4^2)$\vspace{0.05cm}\\\vspace{0.05cm}
$\sigma \sim \mathrm{Exponential}(1)$\vspace{0.5cm}\\\vspace{0.05cm}
       $\log_{10} M_\mathrm{BH} \sim \mathcal{N}(0,2^2)$\vspace{0.05cm}\\\vspace{0.05cm}
        $\log_{10} L \sim \mathcal{N}(0,2^2)$
        \vspace{0.05cm}\\\vspace{0.05cm}
        $\log_{10} M_\mathrm{BH}^\mathrm{obs} \sim \mathcal{N}\left((\log_{10} M_\mathrm{BH},\sigma_{\log_{10} M_\mathrm{BH}^\mathrm{obs}}^2\right)$
        \vspace{0.05cm}\\\vspace{0.05cm}
        $\log_{10} L^\mathrm{obs}  \sim \mathcal{N}\left(\log_{10} L,\sigma_{\log_{10} L^\mathrm{obs}}^2\right)$
        \vspace{0.05cm}\\\vspace{0.05cm}
        $\log_{10} t_\mathrm{b}^\mathrm{true} \sim \mathcal{N}\left(\log_{10} t_\mathrm{b},\sigma^2\right)$
        \vspace{0.05cm}\\\vspace{0.05cm}

        $\log_{10} t_\mathrm{b}^\mathrm{obs} \sim \mathcal{N}\left(\log_{10} t_\mathrm{b}^\mathrm{true},\sigma_{\log_{10} t_\mathrm{b}^\mathrm{obs}}^2\right)$};

        \end{tikzpicture}
\caption{Directed acyclic diagram representing the error-in-variable modelling of \cref{eq:mchardyrelation}. The parameters are shown in orange circles and the data/observation quantities are shown in purple dashed circles. Quantities in the rectangle are different for each source in the sample.  The prior distributions and relations between parameters are presented on the right. }
\label{fig:dag}
\end{figure*}

\onecolumn

\clearpage

\renewcommand{\arraystretch}{1.25}\begin{longtable}{*{8}{l}}\caption{Physical properties of the objects studied. The columns are: (1) Object name, (2) redshift obtained from SIMBAD \citep{2000A&AS..143....9W}, (3) type from \citet{2010A&A...518A..10V}, $^\dagger$ indicates that the type has been adjusted to match the X-ray literature, (4) recalibrated black hole mass in units of $10^7 M_{\odot}$, more details on the recalibration are available in \cref{apdx:calibmasses}, (5) method used to estimate the mass, RM indicates to reverberation mapping and SR indicates scaling relation, (6) decimal logarithm of the continuum $5100$\textup{~\AA} luminosity in units of erg s$^{-1}$, (7) reference(s) and (8) square root of 
the rms amplitude of variability (expressed in percent) computed using segments of 30\,ks as explained in \cref{sec:results}. References: {\protect\hypertarget{bib_1}{1}} {\citet{2013ApJ...769..128B}}, {\protect\hypertarget{bib_2}{2}} {\citet{2011ApJ...732..121B}}, {\protect\hypertarget{bib_3}{3}} {\citet{2021ApJ...906...50B}}, {\protect\hypertarget{bib_4}{4}} {\citet{2023ApJ...959...25B}}, {\protect\hypertarget{bib_5}{5}} {\citet{2009ApJ...705..199B}}, {\protect\hypertarget{bib_6}{6}} {\citet{2007ApJ...662..205B}}, {\protect\hypertarget{bib_7}{7}} {\citet{2016ApJ...830..136B}}, {\protect\hypertarget{bib_8}{8}} {\citet{2006ApJ...651..775B}}, {\protect\hypertarget{bib_9}{9}} {\citet{2013ApJ...767..149B}}, {\protect\hypertarget{bib_10}{10}} {\citet{2016A&A...591A..88B}}, {\protect\hypertarget{bib_11}{11}} {\citet{2015A&A...578A..28B}}, {\protect\hypertarget{bib_12}{12}} {\citet{1993A&A...279...53B}}, {\protect\hypertarget{bib_13}{13}} {\citet{2020ApJ...905...77B}}, {\protect\hypertarget{bib_14}{14}} {\citet{2018A&A...615A.167C}}, {\protect\hypertarget{bib_15}{15}} {\citet{1998ApJ...500..162C}}, {\protect\hypertarget{bib_16}{16}} {\citet{2018ApJ...866..133D}}, {\protect\hypertarget{bib_17}{17}} {\citet{2009ApJ...702.1353D}}, {\protect\hypertarget{bib_18}{18}} {\citet{2010ApJ...721..715D}}, {\protect\hypertarget{bib_19}{19}} {\citet{2006ApJ...653..152D}}, {\protect\hypertarget{bib_20}{20}} {\citet{2015ApJ...806...22D}}, {\protect\hypertarget{bib_21}{21}} {\citet{2016ApJ...825..126D}}, {\protect\hypertarget{bib_22}{22}} {\citet{2017ApJ...840...97F}}, {\protect\hypertarget{bib_23}{23}} {\citet{2021ApJ...912...92F}}, {\protect\hypertarget{bib_24}{24}} {\citet{2012ApJ...755...60G}}, {\protect\hypertarget{bib_25}{25}} {\citet{2004AJ....127..156G}}, {\protect\hypertarget{bib_26}{26}} {\citet{2015ApJ...804..138H}}, {\protect\hypertarget{bib_27}{27}} {\citet{2012MNRAS.420.1825J}}, {\protect\hypertarget{bib_28}{28}} {\citet{1999ApJS..125..317L}}, {\protect\hypertarget{bib_29}{29}} {\citet{2018ApJ...869..137L}}, {\protect\hypertarget{bib_30}{30}} {\citet{2021ApJ...918...50L}}, {\protect\hypertarget{bib_31}{31}} {\citet{2022ApJS..261....5M}}, {\protect\hypertarget{bib_32}{32}} {\citet{2002ApJ...572..746O}}, {\protect\hypertarget{bib_33}{33}} {\citet{2012ApJ...747...30P}}, {\protect\hypertarget{bib_34}{34}} {\citet{2017ApJ...837..131P}}, {\protect\hypertarget{bib_35}{35}} {\citet{2002ApJ...581..197P}}, {\protect\hypertarget{bib_36}{36}} {\citet{2014ApJ...795..149P}}, {\protect\hypertarget{bib_37}{37}} {\citet{2004ApJ...613..682P}}, {\protect\hypertarget{bib_38}{38}} {\citet{1998ApJ...501...82P}}, {\protect\hypertarget{bib_39}{39}} {\citet{2020ApJS..249...17R}}, {\protect\hypertarget{bib_40}{40}} {\citet{2004ApJ...602..635R}}, {\protect\hypertarget{bib_41}{41}} {\citet{2010MNRAS.403.1246S}}, {\protect\hypertarget{bib_42}{42}} {\citet{2022ApJ...925...52U}}, {\protect\hypertarget{bib_43}{43}} {\citet{2016ApJ...824..149W}}.} \label{tab:long} \\

\toprule\toprule
Object  & Redshift & Type & $M_\mathrm{BH}~(10^7 M_\odot)$ & Method & $\log_{10}(\lambda L_\mathrm{5100})$ & Reference(s) & $F_\text{var}\,(\%)$ \\ (1)&(2) & (3)  & (4) & (5) & (6) & (7) & (8)\\ \midrule \endfirsthead

\multicolumn{3}{c}%
{{\bfseries \tablename\ \thetable{} -- continued}} \\
\toprule\toprule
Object  & Redshift & Type & $M_\mathrm{BH}~(10^7 M_\odot)$ & Method &  $\log_{10}(\lambda L_\mathrm{5100})$ & Reference(s) & $F_\text{var}\,(\%)$ \\ (1)& (2) & (3)  & (4) & (5) & (6) & (7) & (8) \\ \midrule \endhead1H 0323+342 & 0.063 & NLS1 & $0.28\pm0.094$ & RM & 43.88 & \protect\hyperlink{bib_1}{43} &$10\pm0.5$   \\ 
1H 0419-577 & 0.10 & S1.5 & $11\pm3.0$ & SR & 44.90 & \protect\hyperlink{bib_2}{25}, \protect\hyperlink{bib_3}{31} &$1.6\pm1$   \\ 
1H 0707-495 & 0.041 & NLS1 & $0.26\pm0.11$ & SR & 43.86 & \protect\hyperlink{bib_4}{14} &$35\pm1$   \\ 
1H 1934-063 & 0.010 & S1.5 & $0.061\pm0.017$ & SR & 43.02 & \protect\hyperlink{bib_4}{14} &$21\pm0.2$   \\ 
Ark 120 & 0.033 & S1 & $11\pm3.8$ & RM & 43.75 & \protect\hyperlink{bib_5}{38}, \protect\hyperlink{bib_6}{9}, \protect\hyperlink{bib_7}{42} &$1.7\pm0.6$   \\ 
Ark 564 & 0.024 & NLS1$^\dagger$ & $0.13\pm0.060$ & SR & 43.51 & \protect\hyperlink{bib_8}{40} &$17\pm0.1$   \\ 
ESO 141-G55 & 0.037 & S1.2 & $8.9\pm2.3$ & SR & 43.77 & \protect\hyperlink{bib_3}{31} &$2.3\pm0.2$   \\ 
ESO 362-G18 & 0.012 & S1.5 & $1.9\pm0.62$ & SR & 42.83 & \protect\hyperlink{bib_3}{31} &$8.8\pm2$   \\ 
ESO 511-G030 & 0.023 & S1 & $1.5\pm0.44$ & SR & 43.04 & \protect\hyperlink{bib_3}{31} &$3.2\pm1$   \\ 
Fairall 9 & 0.046 & S1.2 & $21\pm7.2$ & RM & 43.92 & \protect\hyperlink{bib_9}{37}, \protect\hyperlink{bib_6}{9} &$1.9\pm0.6$   \\ 
IGRJ21277+5656 & 0.015 & NLS1 & $0.94\pm0.24$ & SR & 43.76 & \protect\hyperlink{bib_3}{31} &$15\pm0.6$   \\ 
IRAS 05078+1626 & 0.017 & S1.5 & $1.9\pm0.56$ & SR & 43.00 & \protect\hyperlink{bib_3}{31} &$3.3\pm0.8$   \\ 
IRAS 09149-6206 & 0.057 & S1 & $27\pm7.7$ & SR & 45.18 & \protect\hyperlink{bib_3}{31} &$4.9\pm3$   \\ 
IRAS 13224-3809 & 0.066 & NLS1 & $0.052\pm0.029$ & SR & 43.25\footnote{\label{myfootnote}$\lambda L_{5100}$ is obtained using a scaling relation  with the H$\beta$ luminosity as explained in \cref{sec:scalingrelation} .} & \protect\hyperlink{bib_10}{12} &$48\pm0.6$   \\ 
IRAS 13349+2438 & 0.11 & NLS1 & $6.8\pm2.0$ & SR & 44.64 & \protect\hyperlink{bib_2}{25} &$8.2\pm0.7$   \\ 
IRAS 17020+4544 & 0.061 & NLS1 & $0.12\pm0.048$ & SR & 43.09\footnotemark[9] & \protect\hyperlink{bib_11}{28}, \protect\hyperlink{bib_12}{11}, \protect\hyperlink{bib_13}{10} &$7.1\pm0.4$   \\ 
IRAS F12397+3333 & 0.044 & NLS1 & $0.83\pm0.42$ & RM & 44.23 & \protect\hyperlink{bib_14}{20} &$8.6\pm0.5$   \\ 
MCG 08-11-011 & 0.020 & S1.5 & $2.9\pm0.88$ & RM & 43.59 & \protect\hyperlink{bib_15}{22} &$2.7\pm0.4$   \\ 
MCG-6-30-15 & 0.0071 & S1.5 & $0.20\pm0.098$ & RM & 41.65 & \protect\hyperlink{bib_16}{7} &$20\pm0.2$   \\ 
MR 2251-178 & 0.017 & NLS1 & $11\pm3.7$ & SR & 44.11 & \protect\hyperlink{bib_3}{31} &$0.91\pm0.9$   \\ 
MS 22549-3712 & 0.043 & S1.5 & $0.37\pm0.11$ & SR & 43.34 & \protect\hyperlink{bib_2}{25} &$12\pm0.6$   \\ 
Mkn 1044 & 0.035 & NLS1 & $0.15\pm0.056$ & RM & 43.10 & \protect\hyperlink{bib_17}{26}, \protect\hyperlink{bib_18}{21} &$20\pm0.2$   \\ 
Mkn 1048 & 0.045 & NLS1 & $1.9\pm2.5$ & RM & 43.98 & \protect\hyperlink{bib_7}{42} &$3.7\pm1$   \\ 
Mkn 110 & 0.089 & S1.2 & $2.9\pm1.1$ & RM & 43.86 & \protect\hyperlink{bib_5}{38}, \protect\hyperlink{bib_7}{42} &$2.5\pm0.3$   \\ 
Mkn 142 & 0.030 & S1 & $0.48\pm0.17$ & RM & 43.56 & \protect\hyperlink{bib_19}{33}, \protect\hyperlink{bib_14}{20}, \protect\hyperlink{bib_20}{29} &$12\pm0.4$   \\ 
Mkn 1501 & 0.026 & NLS1 & $12\pm4.7$ & RM & 44.32 & \protect\hyperlink{bib_21}{24} &$1.9\pm2$   \\ 
Mkn 279 & 0.017 & NLS1 & $2.8\pm1.0$ & RM & 43.64 & \protect\hyperlink{bib_9}{37}, \protect\hyperlink{bib_6}{9} &$2.7\pm0.3$   \\ 
Mkn 335 & 0.078 & NLS1 & $1.4\pm0.52$ & RM & 43.72 & \protect\hyperlink{bib_9}{37}, \protect\hyperlink{bib_6}{9}, \protect\hyperlink{bib_1}{43} &$12\pm2$   \\ 
Mkn 359 & 0.031 & NLS1 & $0.23\pm0.078$ & SR & 43.08 & \protect\hyperlink{bib_22}{41}, \protect\hyperlink{bib_3}{31} &$19\pm0.5$   \\ 
Mkn 478 & 0.035 & S1.5 & $1.3\pm0.45$ & SR & 44.33 & \protect\hyperlink{bib_2}{25}, \protect\hyperlink{bib_22}{41} &$12\pm0.4$   \\ 
Mkn 493 & 0.026 & S1 & $0.17\pm0.051$ & RM & 43.11 & \protect\hyperlink{bib_14}{20} &$15\pm0.6$   \\ 
Mkn 509 & 0.019 & S1.5 & $11\pm2.8$ & RM & 44.13 & \protect\hyperlink{bib_5}{38}, \protect\hyperlink{bib_6}{9} &$1.8\pm0.3$   \\ 
Mkn 590 & 0.013 & NLS1 & $3.9\pm2.0$ & RM & 43.33 & \protect\hyperlink{bib_5}{38}, \protect\hyperlink{bib_6}{9} &$1.6\pm4$   \\ 
Mkn 6 & 0.022 & S1.2 & $12\pm3.2$ & RM & 43.75 & \protect\hyperlink{bib_21}{24} &$8.4\pm4$   \\ 
Mkn 766 & 0.031 & S1.5 & $0.14\pm0.13$ & RM & 42.51 & \protect\hyperlink{bib_23}{5}, \protect\hyperlink{bib_6}{9} &$18\pm0.3$   \\ 
Mkn 79 & 0.037 & S1.5 & $4.2\pm2.1$ & RM & 43.57 & \protect\hyperlink{bib_6}{9}, \protect\hyperlink{bib_24}{13} &$4.7\pm3$   \\ 
Mkn 817 & 0.065 & S1.5 & $6.5\pm2.2$ & RM & 43.68 & \protect\hyperlink{bib_6}{9}, \protect\hyperlink{bib_25}{30} &$5.0\pm3$   \\ 
Mkn 841 & 0.039 & NLS1 & $4.9\pm2.2$ & RM & 43.83 & \protect\hyperlink{bib_7}{42} &$3.6\pm0.5$   \\ 
NGC 1566 & 0.0047 & S1.5 & $0.33\pm0.11$ & SR & 42.64 & \protect\hyperlink{bib_3}{31} &$6.5\pm0.7$   \\ 
NGC 2617 & 0.014 & S1.8 & $2.5\pm0.70$ & RM & 42.87 & \protect\hyperlink{bib_15}{22}, \protect\hyperlink{bib_26}{23} &$3.8\pm0.3$   \\ 
NGC 3227 & 0.0033 & S1.5 & $0.55\pm0.25$ & RM & 42.35 & \protect\hyperlink{bib_27}{18}, \protect\hyperlink{bib_28}{16}, \protect\hyperlink{bib_29}{4} &$16\pm0.9$   \\ 
NGC 3516 & 0.0087 & S1.5 & $2.7\pm1.0$ & RM & 43.23 & \protect\hyperlink{bib_27}{18}, \protect\hyperlink{bib_28}{16} &$7.9\pm1$   \\ 
NGC 3783 & 0.0090 & S1.5 & $2.4\pm0.88$ & RM & 42.78 & \protect\hyperlink{bib_30}{32}, \protect\hyperlink{bib_6}{9}, \protect\hyperlink{bib_31}{3} &$6.7\pm1$   \\ 
NGC 4051 & 0.0020 & NLS1 & $0.097\pm0.043$ & RM & 41.78 & \protect\hyperlink{bib_32}{17}, \protect\hyperlink{bib_15}{22} &$37\pm0.3$   \\ 
NGC 4151 & 0.0032 & S1.5 & $3.1\pm0.90$ & RM & 42.35 & \protect\hyperlink{bib_33}{8}, \protect\hyperlink{bib_6}{9}, \protect\hyperlink{bib_28}{16} &$2.0\pm1$   \\ 
NGC 4593 & 0.0083 & S1 & $0.85\pm0.29$ & RM & 42.87 & \protect\hyperlink{bib_34}{19}, \protect\hyperlink{bib_6}{9}, \protect\hyperlink{bib_35}{1} &$13\pm0.3$   \\ 
NGC 4748 & 0.014 & NLS1 & $0.29\pm0.14$ & RM & 42.49 & \protect\hyperlink{bib_19}{33}, \protect\hyperlink{bib_6}{9} &$18\pm0.2$   \\ 
NGC 5548 & 0.0060 & S1i & $5.4\pm2.0$ & RM & 43.16 & \protect\hyperlink{bib_36}{35}, \protect\hyperlink{bib_37}{6}, \protect\hyperlink{bib_23}{5}, \protect\hyperlink{bib_27}{18}, \protect\hyperlink{bib_19}{33}, \protect\hyperlink{bib_6}{9}, \protect\hyperlink{bib_38}{34} &$3.8\pm2$   \\ 
NGC 6814 & 0.017 & S1.5 & $1.6\pm0.62$ & RM & 42.05 & \protect\hyperlink{bib_19}{33}, \protect\hyperlink{bib_6}{9} &$19\pm0.3$   \\ 
NGC 7213 & 0.0058 & S1.5$^\dagger$ & $2.1\pm0.58$ & SR & 41.66 & \protect\hyperlink{bib_3}{31} &$2.1\pm0.6$   \\ 
NGC 7469 & 0.0048 & S1 LINER & $1.3\pm0.46$ & RM & 43.50 & \protect\hyperlink{bib_39}{15}, \protect\hyperlink{bib_6}{9}, \protect\hyperlink{bib_40}{36}, \protect\hyperlink{bib_25}{30} &$5.7\pm0.2$   \\ 
PG 1448+273 & 0.016 & S1.5 & $0.63\pm0.21$ & SR & 44.25 & \protect\hyperlink{bib_2}{25}, \protect\hyperlink{bib_41}{27}, \protect\hyperlink{bib_42}{39} &$20\pm0.5$   \\ 
RE J1034+396 & 0.064 & NLS1 & $0.11\pm0.050$ & SR & 43.25 & \protect\hyperlink{bib_2}{25}, \protect\hyperlink{bib_22}{41}, \protect\hyperlink{bib_41}{27} &$9.3\pm0.4$   \\ 
RX J0439.6-5311 & 0.043 & NLS1 & $0.22\pm0.11$ & SR & 44.31 & \protect\hyperlink{bib_2}{25} &$21\pm0.5$   \\ 
Zw 229-015 & 0.24 & NLS1 & $0.83\pm0.27$ & RM & 42.85 & \protect\hyperlink{bib_43}{2} &$12\pm0.6$   \\   \midrule
NGC 5506 & 0.0059 & NLS1 & $10$ & $M-\sigma$ & - &  \cref{apdx:calibmasses}  &$5.4\pm 3$\\ 
\bottomrule\bottomrule
\end{longtable}

\afterpage{

\begin{landscape}\renewcommand{\arraystretch}{1.25}\begin{longtable}{*{12}{l}}\caption{Posterior medians and uncertainties of the single- and double-bending power-law model of sources with evidence for a bend. (1) name of the source, (2) low-frequency slope, (3) low-frequency bend in day$^{-1}$, (4) intermediate or high-frequency slope, (5) high-frequency bend, (6) high-frequency slope, (7) variance parameter, (8) scaling parameter on the error--bars,(9) mean of the time series, (10) inter-calibration parameter, (11) natural logarithm of the evidence, (12) natural logarithm of the Bayes factor in factor of the double bending model. The prior distribution is listed under each parameter, $\mathcal{U}$ indicates a uniform distribution and $\mathcal{N}$ indicates a normal distribution. $\bar{x}$ and $s^2$ are respectively the sample mean and variance of a subset of the light curve of duration $T$ and minimum sampling $\Delta t$, as introduced in \citetalias{2025MNRAS.539.1775L}. Uncertainties on the parameters are given with a $68\%$ credible interval computed using the $16^\mathrm{th}$ and $84^\mathrm{th}$ percentiles.}\label{tab:results}\\
    \toprule\toprule
    Source & $\alpha_1$ & $f_{b,1}$  & $\alpha_2$ & $f_{b,2}$  & $\alpha_3$ & variance  & $\nu$ & $\mu$ & $\gamma$ & $\ln Z$ & $\ln\mathcal{B}$ \\
    & $\mathcal{U}\left[0,1.5\right]$ &  $\mathrm{Log}\mathcal{U}\left[\frac{1}{5T},\frac{5}{2\Delta t}\right]$ & $\mathcal{U}\left[\alpha_1,6\right]$ & $\mathrm{Log}\mathcal{U}\left[f_1,\frac{5}{2\Delta t}\right]$ & $\mathcal{U}\left[\alpha_2,6\right]$ & $\mathrm{Log}\mathcal{N}(-3,2)$ & $\mathrm{Gamma}(2,0.5)$ &$\mathcal{N}(\bar{x},\beta s^2)$ & $\mathrm{Log}\mathcal{N}(-0.1,0.2)$ &  & \\  
    (1) & (2) & (3)  & (4) & (5) &  (6)& (7) & (8) & (9) & (10) & (11) & (12)  
    \\ \midrule \endfirsthead

    \multicolumn{3}{c}%
    {{\bfseries \tablename\ \thetable{} -- continued}} \\
    \toprule\toprule
    Source & $\alpha_1$ & $f_{b,1}$  & $\alpha_2$ & $f_{b,2}$  & $\alpha_3$ & variance  & $\nu$ & $\mu$ & $\gamma$ & $\ln Z$ & $\ln\mathcal{B}$ \\
    % &  &  d$^{-1}$ &  & d$^{-1}$ &  & Count$^2$ s$^{-2}$ &  & Count s$^{-1}$ &  &  & \\  
    (1) & (2) & (3)  & (4) & (5) &  (6)& (7) & (8) & (9) & (10) & (11) & (12)  \\ \midrule \endhead Ark 564 & $0.46\pm{0.2}$ & $0.53^{+0.5}_{-0.2}$ & $1.6\pm{0.07}$ & $140\pm{10}$ & $4.6^{+0.6}_{-0.5}$ & $0.11^{+0.02}_{-0.01}$ & $0.75\pm{0.2}$ & $3.1\pm{0.07}$ & $0.93^{+0.07}_{-0.06}$ & $7665.56$& $26$ \\& $1.2\pm{0.04}$ & $60\pm{10}$ & $3.1\pm{0.2}$ & - & - & $0.58^{+0.3}_{-0.2}$ & $0.40\pm{0.1}$ & $3.1\pm{0.5}$ & $0.94\pm{0.07}$ & $7639.38$& \\\midrule
  1H 1934-063 & $0.19^{+0.2}_{-0.1}$ & $0.61^{+0.4}_{-0.3}$ & $1.6^{+0.09}_{-0.08}$ & $110^{+10}_{-20}$ & $4.8^{+0.7}_{-0.6}$ & $0.12^{+0.03}_{-0.02}$ & $0.83\pm{0.1}$ & $2.4^{+0.08}_{-0.07}$ & $0.96\pm{0.05}$ & $3000.22$& $14$ \\& $1.2^{+0.06}_{-0.07}$ & $60\pm{10}$ & $3.5^{+0.4}_{-0.3}$ & - & - & $0.99^{+0.9}_{-0.4}$ & $0.66\pm{0.1}$ & $2.4^{+0.6}_{-0.7}$ & $0.98\pm{0.05}$ & $2985.92$& \\\midrule
  IRAS 13224-3809 & $0.27\pm{0.1}$ & $1.8^{+0.4}_{-0.3}$ & $2.2\pm{0.05}$ & $190\pm{20}$ & $5.6^{+0.3}_{-0.5}$ & $0.64^{+0.07}_{-0.06}$ & $0.99\pm{0.03}$ & $0.22\pm{0.08}$ & $0.98\pm{0.02}$ & $5541.26$& $26$ \\& $0.59\pm{0.1}$ & $3.8^{+0.8}_{-0.7}$ & $2.4\pm{0.04}$ & - & - & $0.67^{+0.1}_{-0.07}$ & $0.90\pm{0.03}$ & $0.24\pm{0.1}$ & $0.98\pm{0.02}$ & $5515.43$& \\\midrule
  MCG-6-30-15 & $0.61\pm{0.1}$ & $1.7\pm{1}$ & $1.9\pm{0.3}$ & $50^{+40}_{-30}$ & $3.5^{+0.7}_{-0.4}$ & $0.16^{+0.02}_{-0.01}$ & $0.98^{+0.06}_{-0.07}$ & $2.2\pm{0.07}$ & $1.0\pm{0.03}$ & $7369.14$& $5.7$ \\& $0.81\pm{0.07}$ & $5.9\pm{1}$ & $2.6^{+0.08}_{-0.07}$ & - & - & $0.19^{+0.03}_{-0.02}$ & $0.83^{+0.05}_{-0.06}$ & $2.2\pm{0.1}$ & $1.0\pm{0.03}$ & $7363.47$& \\\midrule
  Mkn 279 & $0.69\pm{0.4}$ & $0.0057^{+0.02}_{-0.004}$ & $1.8^{+0.2}_{-0.1}$ & $1.4\pm{0.4}$ & $5.2^{+0.6}_{-0.8}$ & $0.24^{+0.2}_{-0.07}$ & $1.0\pm{0.04}$ & $2.0\pm{0.5}$ & $1.2^{+0.4}_{-0.3}$ & $2989.60$& $4.3$ \\& $1.4^{+0.07}_{-0.09}$ & $0.78\pm{0.3}$ & $4.1^{+0.9}_{-0.6}$ & - & - & $1.2^{+0.8}_{-0.5}$ & $1.0\pm{0.04}$ & $2.1\pm{2}$ & $1.4^{+0.5}_{-0.4}$ & $2985.33$& \\\midrule
  Mkn 766 & $0.58\pm{0.2}$ & $0.13^{+0.2}_{-0.08}$ & $1.6\pm{0.1}$ & $55\pm{10}$ & $3.9^{+0.5}_{-0.4}$ & $0.35^{+0.08}_{-0.05}$ & $1.0\pm{0.05}$ & $1.8\pm{0.2}$ & $0.67\pm{0.1}$ & $6358.48$& $12$ \\& $1.2^{+0.05}_{-0.06}$ & $18^{+6}_{-5}$ & $2.9\pm{0.2}$ & - & - & $1.7^{+1}_{-0.6}$ & $0.91\pm{0.05}$ & $1.8\pm{0.9}$ & $0.65^{+0.2}_{-0.1}$ & $6346.21$& \\\midrule
  Mkn 1044 & $1.0^{+0.07}_{-0.09}$ & $6.6^{+6}_{-4}$ & $2.0\pm{0.2}$ & $100^{+10}_{-20}$ & $5.1^{+0.6}_{-0.7}$ & $0.38^{+0.2}_{-0.1}$ & $1.1^{+0.07}_{-0.08}$ & $2.2\pm{0.3}$ & $0.98\pm{0.05}$ & $6524.49$& $8.1$ \\& $1.2\pm{0.05}$ & $35\pm{6}$ & $3.3\pm{0.2}$ & - & - & $0.84^{+0.5}_{-0.3}$ & $0.91^{+0.08}_{-0.09}$ & $2.2\pm{0.6}$ & $0.98^{+0.05}_{-0.04}$ & $6516.41$& \\\midrule
  \midrule 1H 0323+342 & $1.1\pm{0.06}$ & $7.9\pm{2}$ & $3.7^{+0.6}_{-0.5}$ & - & - & $0.34^{+0.2}_{-0.09}$ & $0.99\pm{0.03}$ & $1.1^{+0.4}_{-0.3}$ & $0.95^{+0.06}_{-0.05}$ & $2995.91$& \\\midrule
  1H 0419-577 & $1.2^{+0.2}_{-0.4}$ & $0.17^{+0.3}_{-0.1}$ & $3.1^{+1}_{-0.6}$ & - & - & $0.50^{+0.6}_{-0.2}$ & $0.97\pm{0.03}$ & $1.7\pm{0.5}$ & $0.91\pm{0.07}$ & $3964.70$& \\\midrule
  1H 0707-495 & $1.0\pm{0.04}$ & $14^{+3}_{-2}$ & $2.6^{+0.09}_{-0.08}$ & - & - & $1.1^{+0.3}_{-0.2}$ & $0.91\pm{0.04}$ & $0.052\pm{0.5}$ & $0.95^{+0.2}_{-0.1}$ & $1794.42$& \\\midrule
  Ark 120 & $0.77^{+0.3}_{-0.4}$ & $0.11^{+0.1}_{-0.06}$ & $2.8\pm{0.2}$ & - & - & $0.11^{+0.1}_{-0.04}$ & $1.0\pm{0.02}$ & $2.3\pm{0.2}$ & $1.0\pm{0.04}$ & $10139.8$& \\\midrule
  ESO 141-G55 & $0.83^{+0.2}_{-0.3}$ & $0.19^{+0.2}_{-0.1}$ & $3.1^{+0.5}_{-0.3}$ & - & - & $0.11^{+0.1}_{-0.04}$ & $0.94\pm{0.05}$ & $2.6\pm{0.2}$ & $0.98\pm{0.05}$ & $1788.77$& \\\midrule
  ESO 362-G18 & $1.3^{+0.08}_{-0.1}$ & $2.7\pm{1}$ & $3.7^{+0.9}_{-0.6}$ & - & - & $2.7^{+2}_{-1}$ & $1.0\pm{0.04}$ & $0.19\pm{1}$ & $0.52\pm{0.2}$ & $1391.14$& \\\midrule
  ESO 511-G030 & $1.3\pm{0.08}$ & $0.73\pm{0.3}$ & $3.8^{+0.6}_{-0.5}$ & - & - & $1.0^{+0.6}_{-0.4}$ & $1.1\pm{0.04}$ & $0.94^{+0.9}_{-0.8}$ & $0.98^{+0.06}_{-0.05}$ & $2454.11$& \\\midrule
  Fairall 9 & $1.3\pm{0.07}$ & $0.52\pm{0.1}$ & $4.2^{+0.6}_{-0.4}$ & - & - & $0.36^{+0.2}_{-0.1}$ & $0.96\pm{0.03}$ & $2.0\pm{0.5}$ & $0.80\pm{0.1}$ & $5480.57$& \\\midrule
  IGRJ21277+5656 & $0.84^{+0.2}_{-0.3}$ & $1.3^{+2}_{-0.8}$ & $2.1^{+0.2}_{-0.1}$ & - & - & $0.24^{+0.1}_{-0.06}$ & $0.89\pm{0.04}$ & $0.071\pm{0.2}$ & $1.0^{+0.08}_{-0.07}$ & $2869.96$& \\\midrule
  IRAS 13349+2438 & $0.65^{+0.3}_{-0.4}$ & $0.32^{+0.4}_{-0.2}$ & $2.5\pm{0.2}$ & - & - & $0.30^{+0.2}_{-0.09}$ & $0.97\pm{0.04}$ & $0.45\pm{0.2}$ & $1.0\pm{0.07}$ & $2076.50$& \\\midrule
  IRAS 17020+4544 & $1.1^{+0.1}_{-0.2}$ & $33\pm{20}$ & $2.9^{+1}_{-0.8}$ & - & - & $0.059^{+0.08}_{-0.03}$ & $0.91^{+0.08}_{-0.1}$ & $1.3^{+0.2}_{-0.1}$ & $0.80\pm{0.1}$ & $1916.37$& \\\midrule
  IRAS F12397+3333 & $1.2^{+0.07}_{-0.1}$ & $13^{+5}_{-6}$ & $4.1^{+1}_{-0.9}$ & - & - & $0.54^{+0.4}_{-0.2}$ & $1.1\pm{0.06}$ & $1.5\pm{0.5}$ & $1.3^{+0.3}_{-0.2}$ & $1325.39$& \\\midrule
  MCG 08-11-011 & $1.1^{+0.1}_{-0.2}$ & $0.52^{+0.2}_{-0.3}$ & $3.9\pm{1}$ & - & - & $0.36^{+0.3}_{-0.2}$ & $1.0\pm{0.08}$ & $1.9\pm{0.5}$ & $0.64^{+0.3}_{-0.2}$ & $561.491$& \\\midrule
  MR 2251-178 & $1.3^{+0.1}_{-0.2}$ & $0.058^{+0.05}_{-0.03}$ & $3.4^{+0.6}_{-0.4}$ & - & - & $0.71^{+0.6}_{-0.3}$ & $0.98\pm{0.02}$ & $1.6^{+0.6}_{-0.7}$ & $0.89\pm{0.03}$ & $7840.60$& \\\midrule
  MS 22549-3712 & $0.44\pm{0.3}$ & $1.2^{+1}_{-0.6}$ & $2.4^{+0.4}_{-0.3}$ & - & - & $0.071\pm{0.02}$ & $0.93\pm{0.05}$ & $0.46\pm{0.1}$ & $0.64^{+0.1}_{-0.08}$ & $1178.33$& \\\midrule
  Mkn 1048 & $1.2\pm{0.1}$ & $0.40^{+0.3}_{-0.2}$ & $3.1^{+0.4}_{-0.3}$ & - & - & $1.1^{+0.9}_{-0.4}$ & $1.0\pm{0.03}$ & $0.80^{+0.7}_{-0.8}$ & $1.0\pm{0.05}$ & $3919.20$& \\\midrule
  Mkn 110 & $1.3^{+0.07}_{-0.06}$ & $1.1^{+0.4}_{-0.3}$ & $4.6^{+0.9}_{-0.8}$ & - & - & $0.41^{+0.2}_{-0.1}$ & $1.0\pm{0.04}$ & $2.0^{+0.5}_{-0.4}$ & $1.1\pm{0.04}$ & $2990.81$& \\\midrule
  Mkn 142 & $0.78\pm{0.1}$ & $2.2^{+2}_{-1}$ & $2.4^{+0.3}_{-0.2}$ & - & - & $0.35^{+0.1}_{-0.06}$ & $1.0\pm{0.1}$ & $0.67\pm{0.2}$ & $0.99^{+0.07}_{-0.06}$ & $272.301$& \\\midrule
  Mkn 1501 & $0.42\pm{0.3}$ & $0.016^{+0.007}_{-0.005}$ & $4.0\pm{1}$ & - & - & $0.35^{+0.1}_{-0.08}$ & $1.2^{+0.09}_{-0.08}$ & $0.71\pm{0.3}$ & $1.1^{+0.4}_{-0.3}$ & $483.142$& \\\midrule
  Mkn 335 & $1.3\pm{0.05}$ & $2.6^{+1}_{-0.9}$ & $2.5\pm{0.1}$ & - & - & $3.0^{+1}_{-0.8}$ & $0.96\pm{0.02}$ & $0.28\pm{1}$ & $0.98^{+0.06}_{-0.05}$ & $3623.06$& \\\midrule
  Mkn 359 & $0.83\pm{0.1}$ & $3.8\pm{1}$ & $3.3^{+0.4}_{-0.3}$ & - & - & $0.20^{+0.08}_{-0.04}$ & $0.96\pm{0.03}$ & $0.70\pm{0.2}$ & $0.95^{+0.06}_{-0.05}$ & $2412.33$& \\\midrule
  Mkn 478 & $1.0^{+0.1}_{-0.2}$ & $2.8^{+3}_{-2}$ & $2.7^{+0.4}_{-0.3}$ & - & - & $0.30^{+0.2}_{-0.1}$ & $0.97\pm{0.04}$ & $0.93\pm{0.3}$ & $0.79\pm{0.2}$ & $1992.68$& \\\midrule
  Mkn 493 & $0.97\pm{0.1}$ & $8.2^{+7}_{-4}$ & $2.5^{+0.4}_{-0.3}$ & - & - & $0.22^{+0.1}_{-0.07}$ & $0.96\pm{0.05}$ & $0.47\pm{0.2}$ & $0.98\pm{0.07}$ & $1576.84$& \\\midrule
  Mkn 509 & $1.2^{+0.08}_{-0.09}$ & $0.50\pm{0.2}$ & $3.4^{+0.3}_{-0.2}$ & - & - & $0.26^{+0.2}_{-0.09}$ & $1.0\pm{0.02}$ & $2.7\pm{0.4}$ & $0.96\pm{0.1}$ & $14103.9$& \\\midrule
  Mkn 590 & $1.4^{+0.06}_{-0.08}$ & $0.68\pm{0.3}$ & $4.6\pm{1}$ & - & - & $1.9^{+1}_{-0.8}$ & $1.1\pm{0.04}$ & $0.20^{+1}_{-0.9}$ & $0.72^{+0.09}_{-0.08}$ & $1534.94$& \\\midrule
  Mkn 817 & $1.4\pm{0.06}$ & $0.85^{+0.4}_{-0.3}$ & $3.2^{+0.6}_{-0.4}$ & - & - & $4.1^{+2}_{-1}$ & $1.0^{+0.02}_{-0.03}$ & $0.89\pm{2}$ & $0.94^{+0.06}_{-0.05}$ & $1499.88$& \\\midrule
  Mkn 841 & $1.2\pm{0.08}$ & $0.82^{+0.4}_{-0.3}$ & $3.4^{+0.6}_{-0.4}$ & - & - & $0.50^{+0.3}_{-0.2}$ & $1.0\pm{0.03}$ & $1.5\pm{0.5}$ & $0.93\pm{0.04}$ & $5240.63$& \\\midrule
  NGC 1566 & $1.4\pm{0.05}$ & $5.4\pm{2}$ & $3.3^{+0.4}_{-0.3}$ & - & - & $1.9^{+0.9}_{-0.6}$ & $0.98\pm{0.03}$ & $0.044\pm{1}$ & $1.0\pm{0.05}$ & $3846.12$& \\\midrule
  NGC 2617 & $1.2^{+0.1}_{-0.2}$ & $0.26^{+0.6}_{-0.1}$ & $2.8^{+0.7}_{-0.3}$ & - & - & $1.3^{+1}_{-0.5}$ & $1.1^{+0.07}_{-0.06}$ & $1.6^{+0.8}_{-0.9}$ & $1.2^{+0.4}_{-0.3}$ & $1023.14$& \\\midrule
  NGC 3227 & $1.1^{+0.07}_{-0.08}$ & $2.3^{+1}_{-0.8}$ & $2.5^{+0.1}_{-0.09}$ & - & - & $1.5^{+0.8}_{-0.5}$ & $0.92\pm{0.03}$ & $1.3\pm{0.8}$ & $0.70^{+0.1}_{-0.09}$ & $6471.88$& \\\midrule
  NGC 3516 & $1.3\pm{0.09}$ & $0.55^{+0.5}_{-0.3}$ & $2.4\pm{0.1}$ & - & - & $3.1^{+2}_{-1}$ & $0.98\pm{0.03}$ & $0.52\pm{1}$ & $0.98\pm{0.05}$ & $5212.58$& \\\midrule
  NGC 3783 & $0.83^{+0.2}_{-0.3}$ & $0.10^{+0.1}_{-0.06}$ & $2.3\pm{0.1}$ & - & - & $0.68^{+0.4}_{-0.2}$ & $0.97\pm{0.03}$ & $1.2^{+0.4}_{-0.3}$ & $1.0\pm{0.1}$ & $4373.42$& \\\midrule
  NGC 4051 & $1.1\pm{0.05}$ & $39\pm{7}$ & $2.8^{+0.1}_{-0.09}$ & - & - & $1.9^{+1}_{-0.6}$ & $0.43\pm{0.06}$ & $1.7\pm{0.9}$ & $1.1\pm{0.03}$ & $3374.72$\\\midrule
  NGC 4151 & $0.95\pm{0.1}$ & $0.058^{+0.04}_{-0.02}$ & $2.7^{+0.2}_{-0.1}$ & - & - & $0.19^{+0.09}_{-0.05}$ & $1.0\pm{0.02}$ & $0.67\pm{0.2}$ & $1.0\pm{0.02}$ & $9035.85$& \\\midrule
  NGC 4593 & $1.0\pm{0.07}$ & $1.4^{+0.4}_{-0.3}$ & $3.0^{+0.2}_{-0.1}$ & - & - & $0.73^{+0.3}_{-0.2}$ & $1.0^{+0.04}_{-0.03}$ & $1.6\pm{0.5}$ & $0.99\pm{0.02}$ & $3910.48$\\\midrule
  NGC 4748 & $1.1\pm{0.05}$ & $52^{+9}_{-10}$ & $4.6\pm{0.9}$ & - & - & $0.28^{+0.2}_{-0.09}$ & $0.96\pm{0.1}$ & $1.7\pm{0.4}$ & $1.2^{+0.4}_{-0.3}$ & $565.006$& \\\midrule
  NGC 5548 & $1.2^{+0.08}_{-0.09}$ & $0.21^{+0.1}_{-0.07}$ & $2.8^{+0.2}_{-0.1}$ & - & - & $1.0^{+0.5}_{-0.3}$ & $1.0\pm{0.02}$ & $0.89\pm{0.7}$ & $0.92\pm{0.02}$ & $6885.00$\\\midrule
  NGC 6814 & $1.0\pm{0.07}$ & $1.8^{+0.7}_{-0.5}$ & $2.5^{+0.1}_{-0.09}$ & - & - & $1.1^{+0.5}_{-0.3}$ & $0.90^{+0.05}_{-0.04}$ & $1.3\pm{0.6}$ & $1.0^{+0.06}_{-0.05}$ & $2307.96$& \\\midrule
  NGC 7469 & $1.1\pm{0.06}$ & $1.0\pm{0.2}$ & $3.1\pm{0.1}$ & - & - & $0.32^{+0.1}_{-0.08}$ & $0.98\pm{0.02}$ & $2.2\pm{0.3}$ & $1.0\pm{0.02}$ & $11919.5$& \\\midrule
  PG 1448+273 & $1.2^{+0.07}_{-0.08}$ & $10^{+7}_{-4}$ & $2.7^{+0.3}_{-0.2}$ & - & - & $1.2^{+0.8}_{-0.4}$ & $0.97\pm{0.05}$ & $1.3\pm{0.8}$ & $0.95^{+0.3}_{-0.2}$ & $1468.41$& \\\midrule
  RE J1034+396 & $0.98\pm{0.04}$ & $42^{+4}_{-5}$ & $4.1^{+0.9}_{-0.6}$ & - & - & $0.035^{+0.009}_{-0.007}$ & $1.3\pm{0.04}$ & $1.1^{+0.08}_{-0.09}$ & $0.94\pm{0.05}$ & $9622.11$\\\midrule
  RX J0439.6-5311 & $0.89^{+0.1}_{-0.3}$ & $5.6^{+9}_{-4}$ & $2.2^{+0.4}_{-0.2}$ & - & - & $0.20^{+0.1}_{-0.06}$ & $0.91\pm{0.1}$ & $0.44\pm{0.3}$ & $0.73\pm{0.2}$ & $871.643$& \\\midrule
  Zw 229-015 & $1.1\pm{0.1}$ & $1.8^{+0.9}_{-1}$ & $4.2\pm{1}$ & - & - & $1.1^{+0.9}_{-0.4}$ & $0.97\pm{0.1}$ & $0.51\pm{0.7}$ & $1.3^{+0.4}_{-0.3}$ & $143.151$&\\\midrule\bottomrule\bottomrule\end{longtable}\end{landscape}
}
% Don't change these lines
\bsp	% typesetting comment
\label{lastpage}

\end{document}